\documentclass[acmsmall]{acmart}

\usepackage{annotate-equations}
\usepackage{algorithm}
\usepackage{algorithmic}
\usepackage{tcolorbox}
\usepackage{multirow}
\usepackage{array}
\usepackage{epsfig}
\usepackage{natbib}
\usepackage{tabularx}
\usepackage{algorithm}
\usepackage{mathtools}

\usepackage{wrapfig}
\long\def\comment#1{}
\usepackage{multirow}
\usepackage{wrapfig}
\usepackage{textcomp,booktabs}
\usepackage{colortbl}
\usepackage{multirow}
\usepackage{rotating}
\usepackage{epstopdf}
\usepackage{xspace}
\usepackage{bbm}
\definecolor{Gray}{RGB}{230,230,230}
\usepackage{bbold}
\usepackage{graphicx}
\usepackage{caption}
\usepackage{enumitem}
\usepackage{lipsum}
\usepackage{soul}
\usepackage{color}
\usepackage[normalem]{ulem}
\usepackage[usestackEOL]{stackengine}

\newcommand{\todo}[1]{\textcolor{black}{#1}}

\usepackage[labelfont=bf,textfont=normalfont,textfont=sl]{caption}
\usepackage{hyperref}
\hypersetup{
 colorlinks=true,      
 linkcolor=blue,       
 citecolor=magenta,    
 filecolor=cyan,       
 urlcolor=red          
}
\usepackage{balance}
\usepackage{makecell}
\usepackage{sistyle}
\SIthousandsep{,}
\usepackage{filecontents}
\usepackage{tikz}

\usepackage[T1]{fontenc}
\usepackage{mhchem}

\newcommand{\sol}{{ForgetMeNot}}

\newtcolorbox{highlighted}{colback=yellow,coltext=black,breakable}
\definecolor{red_color}{rgb}{1.0, 0.0, 0.0}
\definecolor{black_color}{rgb}{0.0, 0.0, 0.0}
\definecolor{blue_color}{rgb}{0.0, 0.0, 1.0}

\newtcolorbox{myregbox}[3][]
{
  colback  = gray!5, 
  colframe = gray!75, 
  boxsep   = -0.5mm,
  #1,
}

\makeatletter
\newcommand{\removelatexerror}{\let\@latex@error\@gobble}
\makeatother

\begin{document}

\title[\sol{}: Impact of Forever Chemicals Toward Sustainable Large-Scale Computing]{\sol{}: Understanding and Modeling the Impact of Forever Chemicals Toward Sustainable Large-Scale Computing}



\settopmatter{printacmref=false} 
\renewcommand\footnotetextcopyrightpermission[1]{} 
\acmConference{}{}{}
\acmYear{} \acmMonth{}
\acmDOI{} \acmISBN{}

\author{Rohan Basu Roy}
\email{rohanbasuroy@sci.utah.edu}
\affiliation{
    \institution{University of Utah}
    \country{Salt Lake City, USA} 
}

\author{Raghavendra Kanakagiri}
\email{raghavendra@iittp.ac.in}
\affiliation{
    \institution{Indian Institute of Technology Tirupati}
    \country{Chindepalle, India} 
}

\author{Yankai Jiang}
\email{jiang.yank@northeastern.edu}
\affiliation{
    \institution{Northeastern University}
    \country{Boston, USA} 
}

\author{Devesh Tiwari}
\email{d.tiwari@northeastern.edu}
\affiliation{
    \institution{Northeastern University}
    \country{Boston, USA} 
}

\begin{abstract}
\noindent\textbf{Abstract.} \textit{Fluorinated compounds, often referred to as forever chemicals, are critical in various steps of semiconductor fabrication like lithography, etching, chamber cleaning, and others. Forever chemical emissions can exhibit global warming potentials thousands of times greater than carbon dioxide and persist in the atmosphere for millennia. Despite their severe impact, most sustainability works in computer systems have focused on carbon emissions alone. We address this gap by introducing \sol{}, a modeling tool that quantifies fluorinated-compound emissions by integrating fabrication facility-specific practices and hardware specifications, and validate its accuracy using real-world emission data from fabrication facilities. We show how \sol{} can enable fabrication facilities to optimize design and material usage decisions for emission reduction and provide researchers with a methodology to calibrate emission estimates for hardware designs. When \sol{} is applied to analyze emissions for manufacturing CPUs, DRAM, and storage, it illustrates how hardware generations, lithography techniques, and capacities impact fluorinated compound emissions. Finally, we demonstrate how datacenter operators can assemble low-emission servers while balancing performance demands. By factoring in fluorinated emissions into manufacturing decisions, \sol{} paves the way for building more sustainable systems.}
\end{abstract}



\maketitle

\thispagestyle{fancy}
\fancyhf{} 
\fancyfoot[R]{\scriptsize This work has been accepted at ACM SIGMETRICS 2025, Stony Brook, New York, USA}

\section{Introduction}
\label{sec:intro}

\begin{figure}[t]
    \centering
    \includegraphics[scale=0.59]{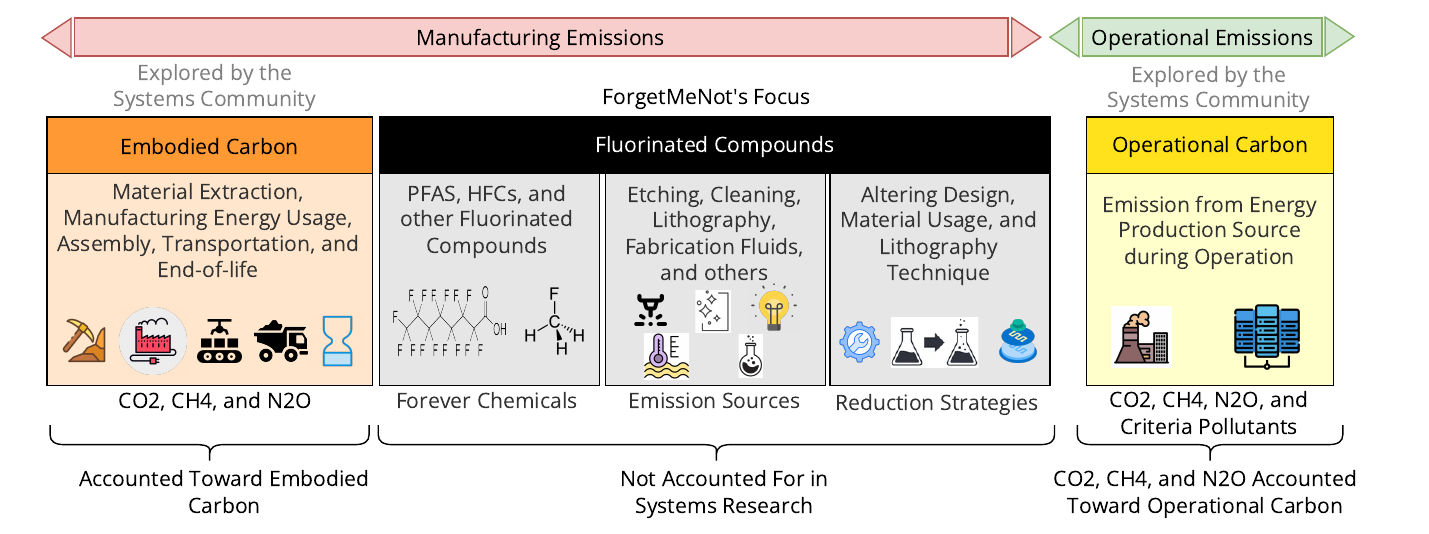}
    \vspace{-4mm}
    \caption{\todo{Systems community has primarily focused on embodied and operational carbon emissions. Fluorinated compound emissions remain unexplored. \sol{} models the fluorinated compound emissions from various sources across the entire semiconductor manufacturing pipeline to provide insights and strategies to reduce overall $CO_2$-eq manufacturing emissions.}}
    \vspace{-4mm}
    \label{fig:intro_cartoon}
\end{figure}


\noindent\textbf{Why should the computer systems community focus on forever chemicals?} The usage of fluorinated compounds including PFAS, hydrofluorocarbons, and others -- often called \emph{forever chemicals} due to their longevity and high resilience to degradation in the environment, are integral ingredients in different steps of modern semiconductor manufacturing like wafer fabrication and photolithography to plasma etching, chamber cleaning, and others~\cite{garg2020review,epaPFASExplained2023,chen2019release,njdohNitrogenTrifluoride2001}. Their global warming potential (GWP) vastly exceeds that of carbon dioxide, with magnitudes up to tens of thousands of times higher. Additionally, these compounds persist in the atmosphere for centuries or millennia, making any emissions effectively permanent contributors to global warming~\cite{ghgprotocolGWPValues2024}.

\todo{The total global warming impact of all greenhouse gases, including carbon dioxide ($CO_2$) and fluorinated compounds, is expressed in terms of $CO_2$-equivalent ($CO_2$-eq), which is the amount of $CO_2$ with the same global warming effect. The unit for expressing the impact of both carbon dioxide ($CO_2$) and fluorinated compounds is grams of $CO_2$-equivalent (g$CO_2$-eq).} \todo{The sustainability studies in the systems and architecture community~\cite{gupta2021chasing,gupta2022act, li2023toward, souza2023ecovisor,bashir2023sustainable}, have primarily focused on the impact of carbon dioxide ($CO_2$) toward embodied carbon (the emissions incurred to produce energy for hardware manufacturing, transportation, and end-of-life processing) and operational carbon (the emissions caused to produce energy during the hardware’s runtime).} \todo{Current sustainability modeling tools (e.g., ACT~\cite{gupta2022act}) primarily focus on estimating carbon dioxide ($CO_2$) emissions and do not account for fluorinated compound emissions toward the overall manufacturing $CO_2$-eq emissions. However, recent findings by Elgamal et al.~\cite{elgamalenvironmental} emphasize that fluorinated compounds can significantly affect the manufacturing carbon footprint. This highlights the need for fluorinated compound modeling in addition to existing embodied carbon modeling tools like ACT (primarily models for only $CO_2$), for a comprehensive assessment of manufacturing $CO_2$-eq emissions.}

\todo{Fig.~\ref{fig:intro_cartoon} shows this point. The systems community has explored the operational emissions and embodied carbon (primarily carbon dioxide emissions) part of the manufacturing emissions. The impact of fluorinated compounds in the manufacturing emissions of computing systems (contributing toward the total $CO_2$-eq emissions), remains unexplored.} Fluorinated compound emissions do not correlate directly with the power grid's energy consumption but arise from \emph{material usage} in specific fabrication processes such as during chamber cleaning, plasma etching, photolithography, usage of heat transfer fluids, and others.  Tracking them demands a detailed understanding of fabrication-specific practices, chemical usage profiles, gas recovery factors, and manufacturing steps at varying technology node sizes. Consequently, although these emissions dwarf carbon emissions in terms of atmospheric impact, they remain largely overlooked in systems research~\cite{elgamalenvironmental}.


\vspace{2mm}

\noindent\textbf{\sol{} addresses a critical research gap.} In this paper, we close this research gap by introducing a modeling tool, \sol{}, that enables the systems community to characterize and mitigate fluorinated-compound emissions in computing hardware manufacturing. To build this tool, we first identify the fluorinated compound emission sources across the various steps of a semiconductor manufacturing pipeline: from wafer fabrication to final packaging. Based on this, we build \sol{} which incorporates fabrication facility-specific chemical usage practices, node-specific lithography details (e.g., DUV v/s EUV), and hardware-specific attributes (e.g., number of cores, die area, TDP, wafer yield, nanometer scale factor, etc.) to estimate the net fluorinated compound releases for new hardware designs. We validate \sol{} against real fab emissions and show that our model captures emissions within 5\% of measured values across facilities. 

\vspace{2mm}

\noindent\textbf{\sol{} can be used to make actionable manufacturing decisions, hardware emission analysis, and building sustainable servers.} \sol{} enables informed design decisions and provides fabrication facilities with a tool to model emissions based on material usage, helping to minimize forever-chemical emissions without compromising performance or yield targets. For instance, we demonstrate how \sol{} can model strategies to reduce emissions, such as adjusting core counts and cache sizes to minimize die area (and associated fluorinated compound usage), adopting advanced lithography techniques with fewer etch steps, optimizing manufacturing processes (e.g., reducing high-GWP fluorinated compound usage during chamber-clean frequencies with more number of etching steps), and implementing advanced gas-capture and recycling technologies. \todo{Since \sol{} models fluorinated compound emissions from each source across the semiconductor manufacturing pipeline, it provides emissions estimates that can be used to explore emission reduction strategies across different manufacturing steps. These capabilities are not supported by the embodied carbon-centric models (account for carbon dioxide emissions), currently used in the computer systems community (\textit{e.g.},~\cite{gupta2022act}), as they do not capture fluorinated compound emissions and their distribution across manufacturing steps.}


For the broader systems community's benefit, we design a methodology to show how public data sources (e.g., Toxic Release Inventory logs~\cite{epaToolbox}) can be used to calibrate \sol{} for specific fabs, enabling consistent carbon-plus-fluorine emission estimates for proposed hardware designs.  Using \sol{}, we analyze fluorinated emissions across hardware types, highlighting trade-offs between performance and sustainability, driven by fabrication methods and technology transitions. We show how datacenter operators can use \sol{} to select hardware configurations that minimize emissions while maintaining competitive performance. \todo{Furthermore, we show that hardware determined as most sustainable when considering only embodied carbon -- using existing modeling tools (e.g., ~\cite{gupta2022act}) that primarily account for carbon dioxide -- differs from the hardware choices determined when accounting for both embodied carbon and fluorinated compounds (via \sol{}). Modeling and including the fluorinated compound emissions in the total $CO_2$-eq emission estimation provides a more accurate representation of sustainable hardware.}

\vspace{2mm}

\noindent\textbf{Contributions.} Specifically, this paper makes the following key contributions:

\vspace{1mm}

\noindent 1. \textit{Reveal the hidden impact of forever chemicals.} We show the impact on global warming due to fluorinated compound emissions during hardware manufacturing, highlighting the need to consider their effect while making manufacturing, fabrication, and design decisions. 

\vspace{1mm}

\noindent 2. \textit{Introduce ForgetMeNot modeling tool.} We propose \emph{ForgetMeNot}, a tool integrating fabrication facility-specific practices, hardware parameters, and specifications to model fluorinated compound emissions from different sources during fabrication. 

\vspace{1mm}

\noindent 3. \textit{Show how \sol{} can help fabrication facilities and researchers.} We show how fabrication facilities can use \sol{} to make design and material usage-based decisions to reduce emissions. We design a methodology using public data to calibrate \sol{} for specific fabs, enabling researchers to obtain emission estimates for hardware designs.

\vspace{1mm}

\noindent 4. \textit{Perform hardware emissions analysis.} We use \sol{} to analyze different hardware, showing how node size, lithography, and architecture affect fluorinated compound footprints. 

\vspace{1mm}

\noindent 5. \textit{Design sustainable datacenter servers.} We demonstrate how datacenter operators can assemble hardware that reduces fluorinated emissions during manufacturing. 

\vspace{2mm}


\noindent We have open-sourced \sol{} at: \href{https://doi.org/10.5281/zenodo.15123079}{https://doi.org/10.5281/zenodo.15123079}. We hope this tool will help progress research in this area, by enabling more accurate modeling and analysis of emissions. \todo{We acknowledge that accurately modeling fluorinated emissions is ultimately limited by the quality of the available data and details. We are hopeful that in the future, a single, detailed standardized reporting structure and methodology for fluorinated compounds will make the modeling more accurate and standardized -- \sol{} is the first step toward that goal.} Next, we discuss the impact of fluorinated compound emissions on global warming.  
\section{Impact of fluorinated compound emissions from semiconductor manufacturing}
\label{sec:fluorine_motive}

\begin{figure}[t]
    \centering
    \includegraphics[scale=0.38]{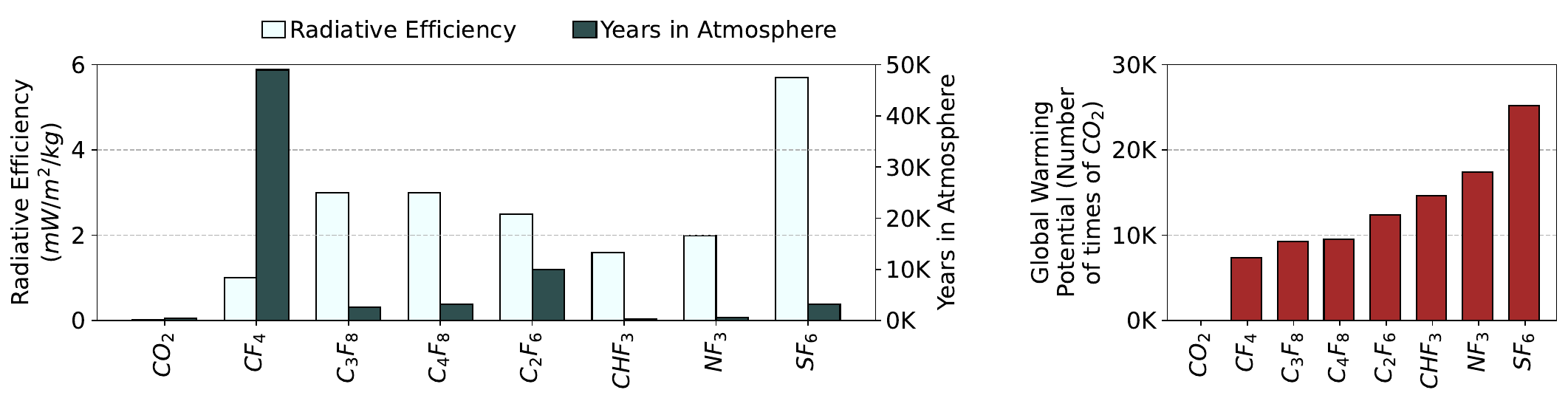}
    \vspace{-4mm}
    \caption{The global warming potential depends on the radiative efficiency and atmospheric lifespan. Fluorinated compounds have tens of thousands of times more impact on global warming than carbon \todo{(obtained from IPCC Global Warming Potential values~\cite{ghgprotocolGWPValues2024})}.}
    \label{fig:global_warming_potential}
    \vspace{-4mm}
\end{figure}

Fluorinated compounds, essential in semiconductor manufacturing, are persistent environmental pollutants with severe health and climate impacts, representing a largely overlooked sustainability concern compared to the systems community's focus on carbon emissions. There are seven main types of fluorinated compounds used in electronics manufacturing, including perfluoroalkyl substances (PFAS), hydrofluorocarbons (HFCs) and other fluorinated compounds: Carbon Tetrafluoride ($\ce{CF4}$), Octafluoropropane ($\ce{C3F8}$), Perfluorocyclobutane ($\ce{C4F8}$), Hexafluoroethane ($\ce{C2F6}$), Trifluoromethane ($\ce{CHF3}$), Nitrogen Trifluoride ($\ce{NF3}$), and Sulfur Hexafluoride ($\ce{SF6}$). They pose severe global warming risks. For example, the simplest fluorinated compound, $\ce{CF4}$, is 7,380 times more potent than carbon dioxide in trapping heat in the atmosphere~\cite{ghgprotocolGWPValues2024}. On the extreme end, $\ce{SF6}$ is 25,200 times more potent than carbon dioxide. A common metric used to compare the impacts of different gases on global warming is the Global Warming Potential (GWP), which measures the heat-trapping ability of a gas relative to carbon dioxide over a specific time period. GWP accounts for two factors: radiative efficiency and atmospheric lifetime. Radiative efficiency measures the effectiveness of a gas at absorbing infrared radiation, while atmospheric lifetime measures the time it takes for a gas to be removed from the atmosphere by natural processes. 

\todo{However, GWP is inherently time-dependent and sensitive to the selected time horizon (typically 20, 100, or 500 years), which can significantly influence how short-lived versus long-lived compounds are evaluated. For example, a compound with a shorter atmospheric lifetime but used in larger quantities may appear more favorable under GWP, even if it causes greater warming in the near term. GWP also assumes natural atmospheric removal processes and does not reflect the potential for future technological interventions to actively remove pollutants. Despite these limitations, GWP remains the widely accepted and standardized metric for cross-compound comparison due to its simplicity, consistency with international climate accounting frameworks (e.g., IPCC, GHG Protocol), and its ability to aggregate the climate impact of diverse compounds into a common, policy-relevant scale~\cite{lynch2020demonstrating}.} GWP of a compound is defined as the ratio of its radiative efficiency with respect to $\ce{CO2}$'s radiative efficiency, multiplied by the ratio of the integral of the time-varying radiative efficiency function of the compound compared to the integral of the time-varying radiative efficiency function of $\ce{CO2}$~\cite{ipccClimateChange2001}.

In Fig.~\ref{fig:global_warming_potential} (left), we compare the radiative efficiency and atmospheric lifespan of the seven main fluorinated compounds used in electronics manufacturing. 
Compounds like $\ce{SF6}$ combine high radiative efficiency with long atmospheric lifetimes, making them particularly potent greenhouse gases. $\ce{CF4}$, despite its lower radiative efficiency, has a remarkably long lifetime, contributing significantly to its environmental impact over millennia. Combining radiative efficiency and atmospheric lifetime, Fig.~\ref{fig:global_warming_potential} (right) shows that the fluorinated gases have GWPs several thousand times greater than $\ce{CO2}$, highlighting their significant impact even in small quantities.

\begin{figure}[t]
    \centering
    \includegraphics[scale=0.37]{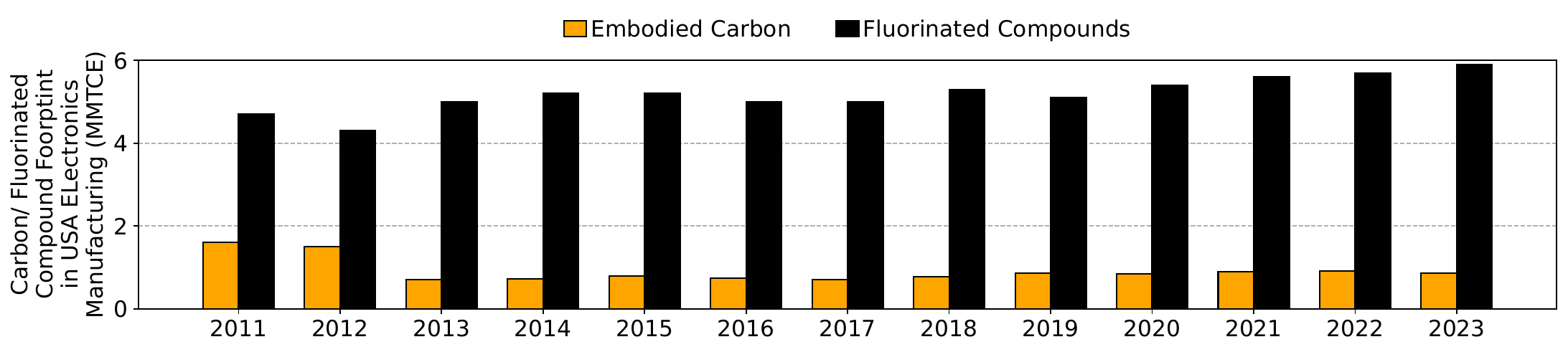}
    \vspace{-4mm}
    \caption{The fluorinated compound footprint is significantly higher than the embodied carbon footprint for semiconductor manufacturing \todo{(obtained from US EPA's GHGRP electronics manufacturing data~\cite{epaGHGRPElectronics})}.}
    \label{fig:carbon_fluorinated_compound_footprint}
    \vspace{-4mm}
\end{figure}

While computing systems research has mostly focused on the carbon footprint, the embodied carbon calculations of computing systems only take into account the $\ce{CO2}$ emissions from energy production during material procurement, manufacturing, and transportation, and do not account for the fluorinated compounds emissions as these are specific to semiconductor fabrication steps and practices followed by fabrication facilities~\cite{gupta2022act,intergovernmental20192019,elgamalenvironmental}.  In Fig.~\ref{fig:carbon_fluorinated_compound_footprint}, we compare the embodied carbon and fluorinated compound footprints within the U.S. electronics manufacturing sector over the past thirteen years~\cite{epaGHGRPElectronics}. The fluorinated compound emissions are the sum of the emissions from seven fluorinated gases commonly used in electronics manufacturing. 
Emissions are measured in Million Metric Tons of Carbon Dioxide Equivalent (MMTCE), a standardized unit used to express the impact of various greenhouse gases (fluorinated compounds) in terms of the amount of carbon dioxide that would have an equivalent global warming effect. MMTCE is calculated as the product of the mass of greenhouse gas in teragrams, its global warming potential, and the factor $\frac{12}{44}$ which is the ratio of carbon's molecular weight to carbon dioxide's. This unified metric allows for a direct comparison between the embodied carbon and fluorinated compound emissions, highlighting the significant impact that fluorinated compound emissions have on the electronics manufacturing industry's total greenhouse gas emissions.

We note that fluorinated compound emissions, aside from in 2010 and 2011, have consistently been 6 to 7 orders of magnitude higher than embodied carbon emissions. These high emission levels are primarily due to the substantial GWP of these compounds and the difficulty in capturing or destroying them. A critical concern is their extremely long atmospheric lifetimes, leading to gradual but significant accumulation in the atmosphere. Fluorinated compound emissions have risen over the years, even as embodied carbon emissions have decreased due to advancements in greener energy and carbon-focused research. This increase is driven by the growing complexity of modern processors and chips, which require finer etching, advanced lithography, and other emission-intensive manufacturing steps. Given their severe environmental impact, research on fluorinated compounds in computing systems is urgently needed. Next, we examine their emission sources across semiconductor manufacturing steps.

\begin{myregbox}{yellow}{} 
\textbf{Takeaway:} While the carbon footprint of computing systems is often the focus of sustainability efforts, this overlooks the significant environmental impact of fluorinated compounds, used extensively in semiconductor manufacturing. Known for their long-lasting presence and potent global warming potential, forever chemicals contribute massively to global warming and environmental toxicity, posing serious unaddressed risks that need immediate attention.
\end{myregbox}

\section{Emission Sources During Manufacturing}
\label{sec:sources}

\noindent Here we examine key semiconductor manufacturing steps and the sources of fluorinated compound emissions. These chemicals are essential in manufacturing for their high reactivity and selectivity~\cite{geng2005semiconductor,ruberti2023chip,osowiecki2024achieving,beu2024overview,dammel2024pfas,bubnova2024road,mann2022semiconductor}.

\vspace{1mm}

\begin{figure*}[t]
    \centering
    \includegraphics[scale=0.32]{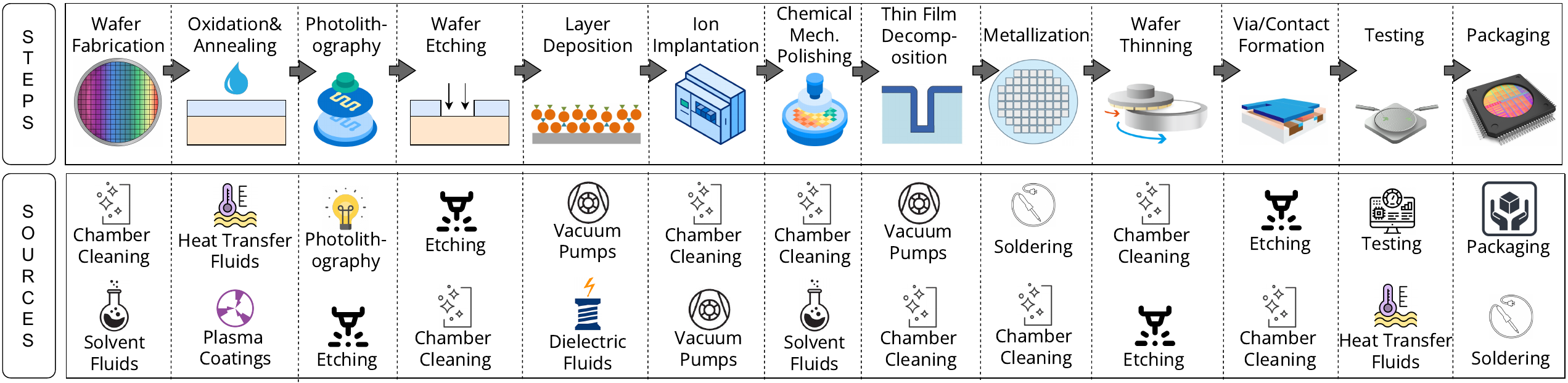}
    \vspace{-4mm}
    \caption{Various steps of manufacturing contribute toward emissions of fluorinated compounds.}
    \label{fig:source}
    \vspace{-4mm}
\end{figure*}

\noindent\textbf{Crystal growth for wafer fabrication.} Wafer fabrication begins with growing pure silicon crystals via the Czochralski or Float Zone methods, producing high-purity ingots that are sliced into wafers. These wafers are polished to achieve a defect-free surface essential for accurate circuit patterning. During this process, plasma cleaning with fluorinated compounds removes residual contaminants from chamber surfaces, ensuring wafer and silicon purity. Following fabrication, wafers are chemically cleaned to eliminate any residual impurities from slicing or polishing. Fluorinated solvents and carrier fluids are used to transport cleaning solutions and remove tough contaminants, further releasing fluorinated compounds.

\vspace{1mm}

\noindent\textbf{Oxidation and annealing.} Thereafter, the wafers undergo oxidation and annealing, which are essential thermal processes that modify wafer properties. Oxidation creates a silicon dioxide insulating layer, while annealing addresses lattice defects, enhancing electrical performance. Both processes are performed in high-temperature furnaces, where Heat Transfer Fluids (HTFs) are used to maintain precise temperature control. HTFs contain fluorinated compounds, leading to emissions. Furthermore, in this step, pulsed-plasma nano-coatings are used to apply ultra-thin layers on the wafers to enhance durability and performance, which are also sources of emissions due to the use of fluorinated compounds in them.

\vspace{1mm}

\noindent\textbf{Photolithography.} Photolithography follows, where a photosensitive polymer (photoresist) is applied to the wafer and exposed to light through a mask, creating circuit patterns. The unneeded portions of the photoresist, which contain fluorinated compounds, are then removed through plasma etching, a process using fluorinated gases for precise material removal without harming the silicon. Plasma cleaning before subsequent steps adds further emissions.

\vspace{1mm}

\noindent\textbf{Etching.} After photolithography, etching further defines circuit patterns by removing materials. Wet etching uses chemicals to dissolve material, and dry etching uses plasma to strip material away with greater control, which is essential for creating fine circuit features.  During etching, fluorinated compounds are used for their chemical reactivity, leading to emissions; they break molecular bonds in the material being etched, forming volatile byproducts that can be easily removed. 

\vspace{1mm}

\noindent\textbf{Layer deposition.} After etching, layers of materials like silicon dioxide, metals, or dielectrics are deposited onto the wafer using Chemical Vapor Deposition (CVD) and Physical Vapor Deposition (PVD). Fluorinated compounds clean vacuum chambers after deposition, preventing residue and causing emissions. Dielectric layers, which insulate metal layers, also use fluorinated compounds for precise etching, contributing further emissions. As circuit patterns develop, photolithography, etching, and deposition steps are repeated, each repetition adding emissions.

\vspace{1mm}

\noindent\textbf{Ion implantation.} In the next step of ion implantation, ions are injected into the wafer to alter the electrical properties of silicon, a critical step to define the behavior of the transistor. While fluorinated compounds aren’t directly involved in implantation, they are used to clean implantation chambers and vacuum pumps, preventing contamination and ensuring process integrity, contributing to indirect emissions.

\vspace{1mm}

\noindent\textbf{Chemical mechanical polishing.} Chemical Mechanical Polishing (CMP) ensures a flat surface after layering, using abrasive slurries with fluorinated solvents and lubricants for a smooth finish, which contributes to emissions. Additionally, chamber cleaning with fluorinated compounds removes polishing byproducts, adding further emissions.

\vspace{1mm}

\noindent\textbf{Thin film decomposition.} To improve circuit durability, protective layers are deposited on the wafers after CMP. Vacuum pumps in conjunction with fluorinated compounds clean deposition chambers (lead to emissions), ensuring new layers adhere correctly and provide the necessary electrical insulation by removing any residuals that could impact performance. 

\vspace{1mm}

\noindent\textbf{Metallization.} Thereafter, the metallization process begins, where metal layers (Copper or Aluminum) are deposited to form connections between transistors. Vapor phase soldering uses fluorinated compounds in inert atmospheres for precision, contributing to emissions as they are vented after each cycle. Fluorinated compounds also clean vacuum chambers during metal deposition.

\vspace{1mm}

\noindent\textbf{Wafer thinning.} To reduce chip height, wafer thinning is performed, followed by plasma cleaning of the wafer's backside, often using fluorinated compounds to eliminate any residual particles. These compounds are used as they are highly effective in cleaning, without causing damage to the delicate circuits on the wafers. This step also involves dry etching to remove excess materials, leading to emissions. 

\vspace{1mm}

\noindent\textbf{Via and contact formation.} After wafer preparation, vias are etched into dielectric layers to connect metal layers. Plasma etching with fluorinated compounds provides the precision needed for this process, emitting fluorinated gases.

\vspace{1mm}

\noindent\textbf{Testing.} Chip testing validates performance, generating significant heat that requires extensive cooling. This requires HTFs, contributing to fluorinated compound emissions.

\vspace{1mm}

\noindent\textbf{Packaging.} Finally, in packaging, the wafer is cut into chips and placed in protective packages with fluorinated compounds in adhesives or coatings, leasing to emissions. Soldering with fluorinated compounds in an inert atmosphere for precision bonds also contributes to emissions.

\begin{figure}[t]
    \centering
    \includegraphics[scale=0.35]{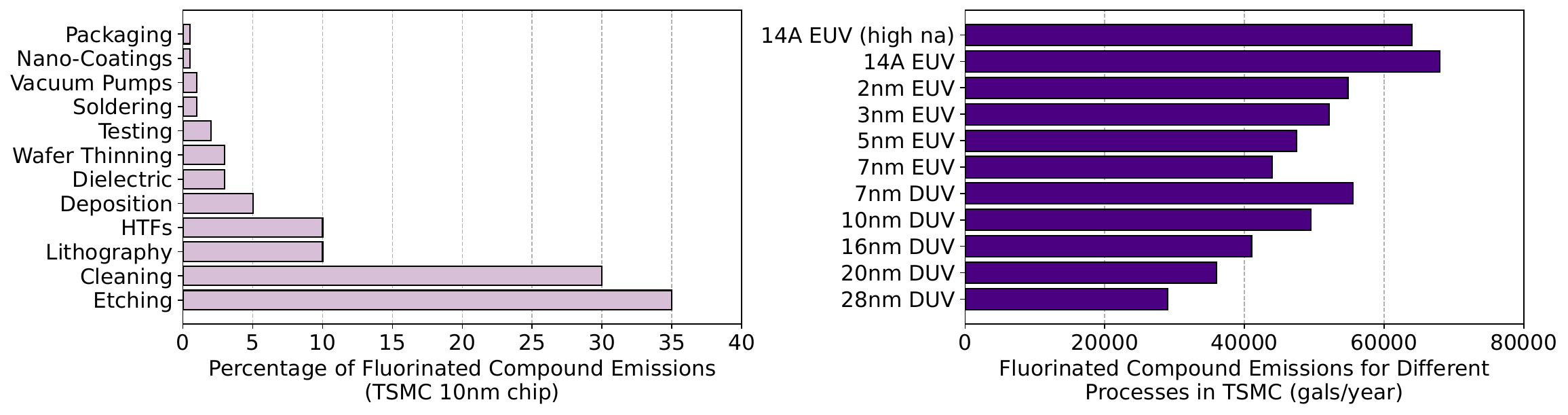}
    \vspace{-4mm}
    \caption{Fluorinated compound emissions vary across emission sources during fabrication and across different nanometer technology node sizes.}
\label{fig:fluorinated_compound_emissions_in_semiconductor_manufacturing}
    \vspace{-4mm}
\end{figure}

Fig.~\ref{fig:source} summarizes the emission sources corresponding to each manufacturing step. We note that certain emission sources like etching, chamber cleaning, heat transfer fluids, and solvent fluids are sources of emission in multiple steps. Additionally, etching and lithography are repeated multiple times during the semiconductor manufacturing pipeline. This leads to more usage and emission of fluorinated compounds from sources like etching, chamber cleaning, lithography, heat transfer fluids, and solvent fluids compared to the other sources (Fig.~\ref{fig:fluorinated_compound_emissions_in_semiconductor_manufacturing} (left)). \todo{The emissions from each source are based on fluorinated compound usage practices and the requirement for manufacturing a 10 nm CPU chip in  TSMC fabrication facilities using DUV lithography~\cite{chemsec,beu2024overview,semiconductors,
iges,epaSemiconductorIndustry}}. 

From Fig.~\ref{fig:fluorinated_compound_emissions_in_semiconductor_manufacturing} (right), we additionally note that as node size (in terms of nanometer technology) reduces, the fluorinated compound emission increases~\cite{techinsightsPFASSemiconductor}. This is because smaller node sizes require more precise and complex manufacturing processes, involving additional etching, lithography, and cleaning steps that utilize fluorinated compounds. As the feature sizes shrink, the need for finer patterning and more frequent cleaning to prevent defects increases, leading to higher usage and emissions of fluorinated compounds. Thus, newer generation hardware (less node size) results in more fluorinated compound emission. Extreme Ultraviolet (EUV) lithography, as opposed to Deep Ultraviolet (DUV) lithography, uses a much shorter wavelength enabling it to pattern smaller features more precisely on wafers. Because EUV lithography requires fewer masks and process steps than DUV, it reduces the use of fluorinated compounds in photolithography and patterning, leading to decreased emissions when transitioning from DUV to EUV (for the same node size). EUV (high na) is an advanced version of EUV lithography that uses a higher numerical aperture (na) lens to further improve resolution and feature sizes, leading to even fewer process steps and etching requirements that further lower fluorinated compound emissions. Next, we discuss the design of \sol{}, which models emissions from each source of fluorinated compound emission in the electronics hardware manufacturing pipeline.

\begin{myregbox}{yellow}{}
\textbf{Takeaway:} The semiconductor manufacturing pipeline emits fluorinated forever chemicals primarily through plasma cleaning, etching, photolithography, and cooling processes, essential for precision but environmentally persistent. Emissions depend on the node size (nanometer technology) and the lithography technique, with newer hardware contributing to larger emissions. Various steps, from fabrication to final testing, contribute to emissions, highlighting the need for targeted emission controls.
\end{myregbox}

\section{\sol{}: Modeling Tool for Forever Chemical Emissions}
\label{sec:model}
\begin{figure}[t]
    \centering
    \includegraphics[scale=0.51]{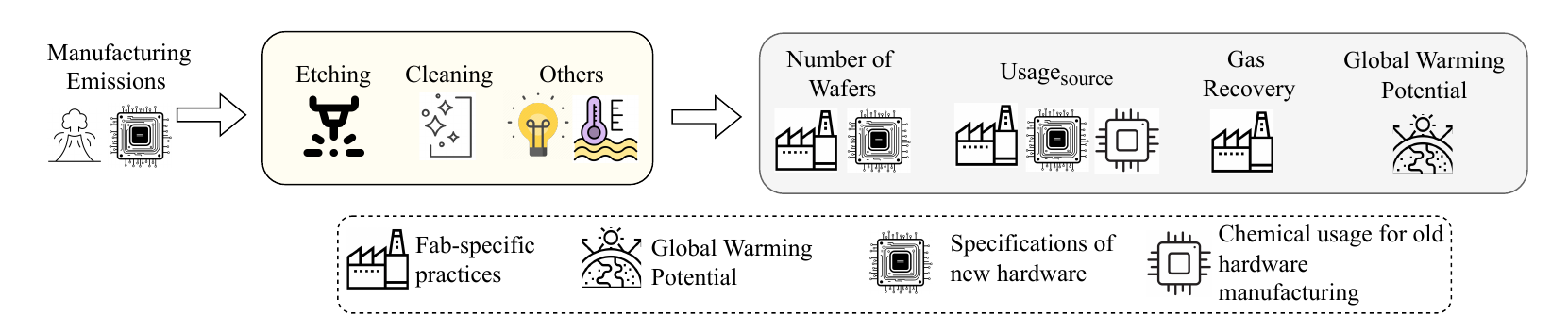}
    \vspace{-4mm}
    \caption{\sol{}'s models fluorinated compound emission of new hardware to be manufactured by considering fabrication facility-specific practices, fluorinated compound usage for an older reference hardware manufacturing, and specifications of the new hardware. }
    \vspace{-4mm}
    \label{fig:model_cartoon}
\end{figure}

\noindent From Sec.~\ref{sec:sources}, we identified that the major sources of fluorinated compound emissions during the manufacturing of server-grade computing systems are the following: etching, plasma chamber cleaning, photolithography, heat transfer fluids, solvent (including carrier and lubricant) fluids, dielectric fluids, wafer thinning, testing, vapor phase soldering, vacuum pumps, pulsed-plasma nanocoatings, and packaging. Based on these emission sources, in this section, we propose \sol{}, a tool for modeling fluorinated compound emissions during the manufacturing of computing systems. \sol{}'s modeling approach leverages fab-specific practices, such as base fluorinated compound usage by sources, along with properties from a reference older generation hardware, to provide detailed emissions estimates for new hardware with given specifications (Fig.~\ref{fig:model_cartoon}). By breaking down emissions from individual sources, \sol{} can equip fabrication facilities with actionable insights to optimize manufacturing steps, reduce specific emissions, and ultimately minimize the overall environmental footprint of next-generation hardware. Following the modeling in Sec.~\ref{sec:model_em},  we show how fabrication facilities can use \sol{} to modify design and manufacturing decisions for emission reduction (Sec.~\ref{sec:interchange}). Thereafter, we discuss the methodology to determine the parameters related to the modeling (Sec.~\ref{sec:methodology}) and validate \sol{}'s accuracy by comparing its estimated emissions against emissions from manufacturing in fabrication facilities (Sec.~\ref{sec:accuracy}).

\subsection{Modeling Emissions from Various Sources}
\label{sec:model_em}
\noindent Emissions from each source depend on the number of wafers processed, the type and quantity of fluorinated compounds used, the gas recovery factor, and the global warming potential (GWP) of the compounds employed in the process~\cite{pelcat2023ghg,ruberti2023chip,illuzzi2010perfluorocompounds}. The gas recovery factor represents the proportion of fluorinated compounds captured and reclaimed during manufacturing processes, thereby directly influencing the net emissions released into the environment. A high gas recovery factor (closer to 1) indicates effective reclamation and lower emissions, while a low factor suggests minimal reclamation and higher emissions. The total emissions (\(E_{\text{Total}}\)) can be expressed as the sum of emissions from each source:

\vspace{-2mm}
\begin{footnotesize}
\begin{equation}
\label{eq:total}
    E_{\text{Total}} = \sum_{i} E_i = \sum_{i} N_{\text{Wafers}} \times \text{Usage}_i \times (1 - \eta_\text{Rec}) \times \text{GWP}_i
\end{equation}
\end{footnotesize}
\vspace{-2mm}

where \(E_i\) represents the emissions from the $i^{th}$ source, \(N_{\text{Wafers}}\) is the number of wafers used to manufacture one hardware component, \(\text{Usage}_i\) denotes the amount of fluorinated compound used per wafer by the emission source \(i\), \(\eta_\text{Rec}\) represents the gas recovery factor, and  \(\text{GWP}_i\) is the Global Warming Potential of the specific fluorinated compound used in emission source \(i\). \(N_{\text{Wafers}}\) depends on the wafer yield factor of the fabrication facility ($\text{Yield}$), and the number of dies that can be created per wafer ($N_{\text{Dies per wafer}}$). The number of dies per wafer depends on the usable wafer area ($A_{\text{Usable wafer}}$), and the area of a die ($A_{\text{Die}} = \pi(\frac{D_{\text{Wafer}}}{2})^{2}$). Again, $A_{\text{Usable wafer}}$ is a product of the area of the wafer ($A_{\text{Wafer}}$) and the wafer usable factor $\gamma_{\text{Usable}}$. Hence, \(N_{\text{Wafers}}\) can be expressed as:

\vspace{-2mm}
\begin{footnotesize}
\begin{align*}
    N_{\text{Wafers}} &= \frac{1}{\text{Yield} \times N_{\text{Dies per wafer}}} 
    = \frac{1}{\text{Yield} \times \frac{A_{\text{Usable wafer}}}{A_{\text{Die}}}} 
    &= \frac{A_{\text{Die}}}{\text{Yield} \times A_{\text{Wafer}} \times \gamma_{\text{Usable}}} = \frac{A_{\text{Die}}}{\text{Yield} \times \pi(\frac{D_{\text{Wafer}}}{2})^{2} \times \gamma_{\text{Usable}}}
\end{align*}
\end{footnotesize}
\vspace{-2mm}

The area of a die ($A_{\text{Die}}$) can be expressed as a function of the features of the hardware components. For example, while manufacturing a CPU core, $A_{\text{Die}} \approx k_{\text{core}}\times N_{\text{core}} + k_{\text{cache}} \times N_{\text{cache}}$, where $N_{\text{core}}$ is the number of CPU cores and $N_{\text{cache}}$ is the size of CPU cache. $k_{\text{core}}$ and $k_{\text{cache}}$ are die scaling parameters for cores and cache, respectively. \todo{We do not enforce integer constraints on $N_{\text{Wafers}}$ and $N_{\text{Dies per wafer}}$; this aligns with industry practices since fabrication facilities inherently account for these constraints in the yield and wafer utilization calculations.}  The parameters defining $N_{\text{Wafers}}$ depend on the type/generation of hardware being manufactured and the specific material usages of the fabrication facilities. For example, for manufacturing $5^{th}$ generation Intel Xeon Platinum CPU with 112 cores, 168 MB cache, produced via 7 nm EUV technology at the Intel fabrication facility in Oregon (Hillsboro, USA), the approximate values of the parameters are $k_{\text{core}} = 4.5$  $\text{mm}^2 / \text{core}$, $k_{\text{cache}} = 0.4$  $\text{mm}^2 / \text{MB}$, $\text{Yield} = 0.8$, $\gamma_{\text{Usable}} = 0.95$, and wafer diameter $D_{\text{wafer}}$ is typically 150mm. Hence, for this processor, $A_{\text{Die}}$ is 571.2$\text{mm}^{2}$, and $N_{\text{Wafers}}$ is calculated to be 0.04255 units. \todo{Wafer usable factor ($\gamma_{\text{Usable}}$) accounts for practical die placement, including spacing and defects. While interleaving dies of different sizes could improve wafer utilization, fabrication facilities typically optimize layouts based on manufacturability and yield constraints.} 

\begin{table*}[t]
\centering
\caption{Global Warming Potential (GWP (gCO$_2$eq)) and fluorinated compound usage models vary for different emission sources.}
\vspace{-2mm}
\label{tab:model}
\footnotesize{
\begin{tabular}{|l|l|l|l|}
\hline
\textbf{Source} & \textbf{Compounds} & \textbf{GWP} & \textbf{Fluorinated Compound Usage Model} \\ \hline
Etching & $CF_4$, $C_2F_6$,$CHF_3$ & 9928 & $\displaystyle \text{Usage}_{\text{Etch}} = k_{\text{Etch}} \times A_{\text{Die}} \times N_{\text{Etch, ref}} \times \left( \dfrac{N_{\text{ref}}}{N} \right)^{\alpha_{\text{Etch}}} \times \phi_{\text{Lith}}$ \\[3mm] \hline
Chamber Cleaning & $NF_3$, $SF_6$ & 19550 & $\displaystyle \text{Usage}_{\text{Clean}} = k_{\text{Clean}} \times A_{\text{Die}} \times N_{\text{Clean, ref}} \times \left( \dfrac{N_{\text{ref}}}{N} \right)^{\alpha_{\text{Clean}}} \times \phi_{\text{Lith}}$ \\[3mm] \hline
Photolithography & $CHF_3$, $C_4F_8$ & 12356 & $\displaystyle \text{Usage}_{\text{Photo}} = k_{\text{Photo}} \times A_{\text{Die}} \times N_{\text{Photo, ref}} \times \left( \dfrac{N_{\text{ref}}}{N} \right)^{\alpha_{\text{Photo}}} \times \phi_{\text{Lith}}$ \\[3mm] \hline
HTFs & $C_3F_8$, $C_4F_8$ & 9405 & $\displaystyle \text{Usage}_{\text{HTF}} = k_{\text{HTF}} \times t_{\text{process, ref}} \times \left( \dfrac{N_{\text{ref}}}{N} \right)^{\alpha_{\text{Time}}} \times \text{TDP}$ \\[3mm] \hline
Solvent Fluids & $C_2F_6$, $NF_3$ & 13140 & $\text{Usage}_{\text{Solv}} = k_{\text{Solv}} \times A_{\text{die}} \times N_{\text{Solv Steps, ref}} \times \left( \frac{N_{\text{ref}}}{N} \right)^{\alpha_{\text{Solv}}}$ \\[3mm] \hline
Dielectric Fluids & $C_4F_8$, $CHF_3$ & 9136 & $\displaystyle \text{Usage}_{\text{Dielec}} = k_{\text{Dielec}} \times A_{\text{Die}}$ \\[3mm] \hline
Wafer Thinning & $SF_6$, $CF_4$ & 17490 & $\displaystyle \text{Usage}_{\text{Thin}} = k_{\text{Thin}} \times A_{\text{wafer}}$ \\[3mm] \hline
Testing & $C_3F_8$, $SF_6$ & 16285 & $\text{Usage}_{\text{Test}} = k_{\text{Test}} \times A_{\text{Die}} \times N_{\text{Test, ref}} \times \left( \frac{N_{\text{ref}}}{N} \right)^{\alpha_{\text{Test}}}$ \\[3mm] \hline
Soldering & $SF_6$ & 17140 & $\displaystyle \text{Usage}_{\text{VPS}} = k_{\text{VPS}} \times N_{\text{Solder, ref}} \times \left( \dfrac{\text{Package Size}}{\text{Package Size}_{\text{ref}}} \right)^{\alpha_{\text{VPS}}}$ \\[3mm] \hline
Vacuum Pumps & $CF_4$, $C_2F_6$ & 9264 & $\displaystyle \text{Usage}_{\text{Vacuum}} = k_{\text{Vacuum}} \times N_{\text{Pump, ref}} \times \left( \frac{N_{\text{ref}}}{N} \right)^{\alpha_{\text{Vacuum}}}$ \\[3mm] \hline
Plasma Coatings & $CHF_3$, $C_4F_8$ & 11000 & $\displaystyle \text{Usage}_{\text{PPNC}} = k_{\text{PPNC}} \times A_{\text{Die}} \times N_{\text{PPNC, ref}} \times \left( \frac{N_{\text{ref}}}{N} \right)^{\alpha_{\text{PPNC}}}$ \\[3mm] \hline
Packaging & $SF_6$, $C_3F_8$ & 18600 & $\displaystyle \text{Usage}_{\text{Pack}} = k_{\text{Pack}} \times \text{Package Size}$ \\[3mm] \hline
\end{tabular}}
\vspace{-4mm}
\end{table*}

Table~\ref{tab:model} shows the Global Warming Potential (GWP) of each source of emissions~\cite{interfaceeuChipProductions,czerniak2018pfc,
epaRiskManagement}. The GWP of emissions from a source is calculated by weighting the GWP of fluorinated compounds used in the source by their emission ratios~\cite{interfaceeuChipProductions}, summing these weighted values, and dividing them by the total of the emission ratios of the fluorinated compounds used in the source. The gas recovery factor ($\eta_{\text{rec}}$) depends on the effectiveness of fabrication facilities in capturing emissions and is usually about $0.9$ for highly advanced facilities, such as those by Intel (Hillsboro, Oregon, USA), where efficient capture and recycling systems are in place to minimize environmental impact. This reflects the proportion of fluorinated compounds captured rather than released, significantly reducing net emissions. To build an accurate emissions model of fluorinated compounds, it is necessary to incorporate process-specific factors like technology node size (nanometer technology used), and lithography technology employed (Extreme Ultraviolet (EUV), or Deep Ultraviolet (DUV)). The \(\text{Usage}_i\) of each emission source $i$ captures these factors, which we discuss next. Throughout the rest of this subsection, following \sol{}'s modeling of emission by each source, we will determine the emissions for a 7 nm, 112-core, 168 MB cache, $5^{th}$ generation Intel Xeon Platinum CPU manufactured via EUV lithography at an Intel fabrication facility at Oregon, based on a 14 nm, $2^{nd}$ generation Intel Xeon Platinum CPU (reference older generation hardware) manufactured via DUV lithography at the same fabrication facility. In Sec.~\ref{sec:methodology}, we discuss the methodology we used to derive the several parameter values used in the modeling.

\vspace{2mm}

\noindent\textbf{Emissions from etching.} The etching process is sensitive to wafer size, node dimensions, and lithography type; finer material removal at smaller technology nodes increases fluorinated compound usage and emissions per unit area. The equation for fluorinated compound usage per wafer in etching ($\text{Usage}_{\text{Etch}}$), as shown in Table~\ref{tab:model}, models these factors. Here, \(k_{\text{Etch}}\) represents the base compound usage coefficient per unit area per etch step at a fabrication facility (typically 0.005 g/mm$^2$/step at Intel fabrication facility in Oregon using DUV lithography). Since the emissions grow with etching steps during manufacturing, the number of etching steps is captured by $N_{\text{Etch, ref}} \times \left( \dfrac{N_{\text{ref}}}{N} \right)^{\alpha_{\text{Etch}}}$, where \(N_{\text{Etch, ref}}\) is the reference number of etching steps at node size (in terms of nanometer technology) $N_{\text{ref}}$ (\textit{e.g.}, 14 nm), and $N$ is the nanometer technology used to manufacture the node for which emission modeling is being performed. $N_{\text{Etch, ref}}$ is $\approx 20$ for manufacturing 14 nm CPU at Intel (Oregon) fabrication facility via DUV lithography (older generation reference hardware). The scaling factor $\alpha_{\text{Etch}}$ ($\approx$ 0.5) models a sub-linear increase in the number of etching steps, acknowledging that etching step requirements rise with smaller nodes but not in direct proportion.  Finally, \(\phi_{\text{Lith}}\) adjusts fluorinated compound usage based on lithography type. EUV (\(\phi_{\text{Lith}} = 0.8\)) requires fewer patterning steps than DUV (\(\phi_{\text{Lith}} = 1\)), typically reducing emissions; the values are based on observed EUV efficiency gains in the number of etching step reduction. 

Combining these values to model $\text{Usage}_{\text{Etch}}$, the total estimated fluorinated compound emissions from etching ($E_{\text{Etch}}$, calculated based on Eq.~\ref{eq:total}) to manufacture a $5^{th}$ generation Intel Xeon Platinum CPU (112 cores, 168 MB cache, produced via 7 nm EUV technology) is 24,560 gCO$_2$eq.

\vspace{2mm}

\noindent\textbf{Emissions from chamber cleaning and photolithography.} Emissions from chamber cleaning and photolithography processes are similarly influenced by node dimensions and lithography technology, as with etching. The fluorinated compound usage per wafer for these processes, \(\text{Usage}_{\text{Clean}}\) and \(\text{Usage}_{\text{Photo}}\), are shown in Table~\ref{tab:model}. Here, \(k_{\text{Clean}}\) and \(k_{\text{Photo}}\) denote the base compound usage coefficients for chamber cleaning and photolithography, respectively (typically \(k_{\text{Clean}}\) = 0.003 g/mm$^2$/step, and \(k_{\text{Photo}}\) = 0.0007 g/mm$^2$/step at Intel fabrication facility in Oregon for manufacturing via DUV lithography). The number of steps in each process is scaled from their reference values, \(N_{\text{Clean, ref}}\) and \(N_{\text{Photo, ref}}\), using the scaling exponents \(\alpha_{\text{Clean}}\) and \(\alpha_{\text{Photo}}\) to adjust for changes in node size from \(N_{\text{ref}}\) to \(N\). \(N_{\text{Clean, ref}}\) and \(N_{\text{Photo, ref}}\) are typically around 15, and 25, respectively for manufacturing a reference older generation (14 nm) hardware. As node nanometer technology decreases, fabricating smaller and more precise features requires additional photolithography steps for accurate patterning and more frequent cleaning to prevent defects, leading to an increase in both photolithography and chamber cleaning steps.
\(\alpha_{\text{Clean}}\) is typically around 0.5, representing a sub-linear increase in cleaning steps due to process optimizations. For photolithography, the scaling factor \(\alpha_{\text{Photo}}\) is usually set to 1, indicating a linear relationship between the reduction in node size and the increase in photolithography steps required for more precise patterning. Similar to calculating $E_{\text{Etch}}$, $E_{\text{Clean}}$ and $E_{\text{Photo}}$ are estimated to be 21,278 gCO$_2$eq and 7568 gCO$_2$eq to manufacture a 112-core $5^{th}$ generation Intel Xeon Platinum CPU at Intel (Oregon) facility. 

\vspace{2mm}

\noindent\textbf{Emissions from heat transfer fluids (HTFs).} Emissions from HTFs are modeled based on the thermal design power (TDP) of the hardware, as higher TDP devices generate more heat during manufacturing, requiring increased use of HTFs for effective cooling (Table~\ref{tab:model}). Unlike etching or lithography, \(A_{\text{Die}}\) and $\phi_{\text{Lith}}$ does not significantly impact HTF usage since thermal management is more related to power dissipation than physical size and lithography process. \(k_{\text{HTF}}\) is the base HTF usage coefficient, representing the amount of HTF required per unit of time per unit of generated heat at a fabrication facility (\(k_{\text{HTF}}\) $\approx 0.0025$ g/hour/W). The processing time impacts emissions proportionally, with longer processes requiring more HTFs. Since processing time increases sub-linearly (the scaling factor \(\alpha_{\text{Time}}\) is typically around 0.5) with a decrease in node size,  $t_{\text{process, ref}} \times \left( \dfrac{N_{\text{ref}}}{N} \right)^{\alpha_{\text{time}}}$ represents the processing time of the newer generation hardware (whose emissions are being determined) compared to the processing time of older generation reference hardware ($t_{\text{process, ref}} \approx$ 20 hours for the 14 nm reference CPU).  Following this modeling, the estimated fluorinated compound emissions from HTFs to manufacture a 112-core $5^{th}$ generation Intel Xeon Platinum CPU (300 W TDP) is 7640 gCO$_2$eq.

\vspace{2mm}

\noindent\textbf{Emissions from Solvent Fluids, Dielectric Fluids, and Wafer Thinning.} Emissions from solvent fluids are modeled based on the die area (\(A_{\text{Die}}\)) and the number of steps where solvents (including carrier and lubricant) fluid is applied (\(N_{\text{Solv Steps, ref}}\)) $\times$ \(\left( \frac{N_{\text{ref}}}{N} \right)^{\alpha_{\text{Solv}}}\), as larger dies and more processing steps require increased use of solvents for cleaning and deposition (Table~\ref{tab:model}). \(N_{\text{Solv Steps, ref}}\) is $\approx$ 10 for the reference 14 nm older generation CPU manufactured at Intel (Oregon) fabrication facility. As node sizes decrease (smaller \(N\)), solvent usage increases due to the need for more precise processing at smaller scales; \(\alpha_{\text{Solv}}\) is typically around 0.5, indicating a sub-linear relationship. \(k_{\text{Solv}}\) is the base solvent usage coefficient and it is $\approx 0.001$ g/mm$^2$/step at Intel fabrication facility in Oregon.
For dielectric fluids, emissions are directly proportional to the die area, as larger dies require more dielectric material for insulation and signal integrity; \(k_{\text{Dielec}}\) is the base dielectric usage coefficient representing the amount used per unit area ($\approx$ 0.01 g/mm$^2$ at Intel fabrication facility). Wafer thinning emissions depend on the wafer area (\(A_{\text{wafer}}\)), with larger wafers requiring more thinning chemicals to achieve the desired thickness; \(k_{\text{Thin}}\) is the base usage coefficient for wafer thinning processes ($\approx$ 0.0002 g/mm$^2$ at Intel fabrication facility). Dielectric fluid use and wafer thinning are uniform processes applied in a single operation, independent of the number of steps. The estimated fluorinated compound emissions from solvent fluids, dielectric fluids, and wafer thinning to manufacture the 112-core Intel CPU are 4064-, 1998-, and 2365 gCO$_2$eq, respectively.

\vspace{2mm}

\noindent\textbf{Emissions from testing.} Emissions from testing processes depend on the die area (\(A_{\text{Die}}\)) and the node size (\(N\)), as smaller node sizes and larger dies require more extensive testing to ensure device reliability (Table~\ref{tab:model}). \(k_{\text{Test}}\) is the base usage coefficient ($\approx$ 0.0001 g/mm$^2$/step at Intel fabrication facility), and \(N_{\text{tests, ref}}\) is the reference number of tests for manufacturing a reference older generation hardware ($\approx$ 20 times for manufacturing 14 nm reference CPU). The scaling term \(\left( \frac{N_{\text{ref}}}{N} \right)^{\alpha_{\text{Test}}}\) captures the increase in testing requirements as node sizes decrease, with \(\alpha_{\text{Test}}\) typically around 1, indicating a linear relationship. This formulation reflects that testing emissions are influenced by the complexity and size of the die rather than factors like TDP or lithography technique, as testing focuses on verifying functionality and detecting defects, which are more prevalent with smaller nodes and larger die areas. Based on the usage modeling, the estimated emissions from testing for manufacturing a 7 nm, 112-core $5^{th}$ generation Intel Xeon Platinum CPU is 1424 gCO$_2$eq. 

\begin{table}[t]
\centering
\caption{\sol{} models emissions during hardware manufacturing based on fabrication facility-specific practices, properties of reference older-generation hardware, and the specifications of the new hardware whose emissions are being modeled. It models emissions from individual sources separately, enabling fine-grained analysis of manufacturing-related emissions.}
\vspace{-2mm}
\label{tab:parameters}
\footnotesize{
\begin{tabular}{|p{0.35\linewidth}|p{0.6\linewidth}|}
\hline
\textbf{Parameter Type} & \textbf{Parameters} \\
\hline
Fabrication facility-specific practices and emission sources & 
\(\eta_{\text{Rec}}\), $\text{Yield}$, $D_{\text{Wafer}}$, $\gamma_{\text{Usable}}$, $A_{\text{Die}}$, $k_{\text{Etch}}$, $k_{\text{Clean}}$, $k_{\text{Photo}}$, $k_{\text{HTF}}$, $k_{\text{Solv}}$, $k_{\text{Dielec}}$, $k_{\text{Thin}}$, $k_{\text{Test}}$, $k_{\text{VPS}}$, $k_{\text{Vacuum}}$, $k_{\text{PPNC}}$, $k_{\text{Pack}}$, $\alpha_{\text{Etch}}$, $\alpha_{\text{Clean}}$, $\alpha_{\text{Photo}}$, $\alpha_{\text{Time}}$, $\alpha_{\text{Solv}}$, $\alpha_{\text{Test}}$, $\alpha_{\text{VPS}}$, $\alpha_{\text{Vacuum}}$, $\alpha_{\text{PPNC}}$, $\phi_{\text{Lith}}$ \\ \hline

Reference older generation hardware & 
$N_{\text{ref}}$, $N_{\text{Etch, ref}}$, $N_{\text{Clean, ref}}$, $N_{\text{Photo, ref}}$, $t_{\text{process, ref}}$, $N_{\text{Solv Steps, ref}}$, $N_{\text{Test, ref}}$, $N_{\text{Solder, ref}}$, $N_{\text{Pump, ref}}$, $\text{Package Size}_{\text{ref}}$ \\ \hline

New hardware specifications & 
$N$, $\text{TDP}$, $\text{Package Size}$, $\text{Cores}$, $\text{Cache}$, $\text{Memory Size}$, $\text{Storage Size}$ \\ \hline
\end{tabular}}
\vspace{-4mm}
\end{table}

\vspace{2mm}

\noindent\textbf{Emissions from vapor phase soldering.} Emissions from soldering processes are modeled based on the package size, as larger packages require more solder joints and soldering steps, leading to increased fluorinated compound usage (Table~\ref{tab:model}). The usage equation includes the base usage coefficient \(k_{\text{VPS}}\) and the reference number of soldering steps, \(N_{\text{Solder, ref}}\) (\(k_{\text{VPS}}\) is 0.4 g/step and \(N_{\text{Solder, ref}}\) $\approx$ 5 for manufacturing 14 nm older generation reference CPU at Intel, Oregon facility). The scaling term \(\left( \frac{\text{Package Size}}{\text{Package Size}_{\text{ref}}} \right)^{\alpha_{\text{VPS}}}\) accounts for the direct relationship of soldering steps with package sizes relative to a reference package size ($\text{Package Size}_{\text{ref}} \approx$ 4000 
 $\text{mm}^2$ for 14 nm reference Intel CPU). \(\alpha_{\text{VPS}}\) is typically around 1, indicating a linear relationship. The estimated emissions from vapor phase soldering for manufacturing a 14 nm, 112-core $5^{th}$ generation Intel Xeon Platinum CPU ($\text{Package Size} \approx$ 2500 $\text{mm}^2$) is 820 gCO$_2$eq. 

\vspace{2mm}

\noindent\textbf{Emissions from vacuum pumps and pulsed-plasma nanocoatings.} Emissions from both vacuum pumps and plasma coatings are influenced by decreasing node sizes, which require higher precision and additional processing steps in semiconductor manufacturing (Table~\ref{tab:model}). For vacuum pumps, the usage equation includes the base usage coefficient \(k_{\text{Vacuum}}\) and the reference number of steps of vacuum pump usage \(N_{\text{Pump, ref}}\) ($k_{\text{Vacuum}} \approx$ 0.02  $\text{g/step}$, and \(N_{\text{Pump, ref}} \approx 50\) for a 14 nm reference CPU at Intel's Oregon facility). The scaling term \(\left( \frac{N_{\text{ref}}}{N} \right)^{\alpha_{\text{Vacuum}}}\) captures the increase in vacuum operations required at smaller node sizes due to the need for cleaner environments and more stringent vacuum conditions; \(\alpha_{\text{Vacuum}}\) is typically around 0.8, indicating a sub-linear relationship. For plasma coatings that are used in the oxidation and annealing steps of manufacturing, the usage equation incorporates the base usage coefficient \(k_{\text{PPNC}}\), the die area \(A_{\text{Die}}\), and the reference number of plasma coating steps \(N_{\text{PPNC, ref}}\) (\(k_{\text{PPNC}} \approx 0.0001 \, \text{g/mm}^2/\text{step}\), and \(N_{\text{PPNC, ref}} \approx 10\) for a 14 nm reference CPU). The scaling term \(\left( \frac{N_{\text{ref}}}{N} \right)^{\alpha_{\text{PPNC}}}\) accounts for the increased need for plasma treatments at smaller node sizes to enhance device performance; \(\alpha_{\text{PPNC}}\) is typically around 1, indicating a linear relationship. Based on this, the emissions from vacuum pumps and plasma coating for manufacturing a 112-core, $5^{th}$ generation Intel Xeon CPU are 618-, and 480 gCO$_2$eq, respectively.

\vspace{2mm}

\noindent\textbf{Emissions from packaging.} Packaging emissions are based on package size, as larger packages require more materials and fluorinated compounds for sealing and protection (Table~\ref{tab:model}). The usage equation includes the base usage coefficient \(k_{\text{Pack}}\) ($\approx$ 0.0002 g/mm$^2$ at Intel's Oregon facility) and the \text{Package Size}. Since packaging is typically done in a single step and is independent of node size or device complexity, emissions scale directly with package size alone. For manufacturing a 7 nm, 112-core Intel Xeon Platinum CPU the packing emission is 356 gCO$_2$eq.

\sol{} estimates the emission from each of the sources with the help of parameters related to fabrication facility-specific practices, reference older generation hardware, and specifications of the new hardware whose emissions are being determined. Table.~\ref{tab:parameters} summarizes all the parameters used in the modeling. Next, we show how fabrication facilities can use \sol{}'s modeling to make design and manufacturing decisions to reduce fluorinated compound emissions. 

\begin{figure*}[t]
    \centering
    \includegraphics[scale=0.37]{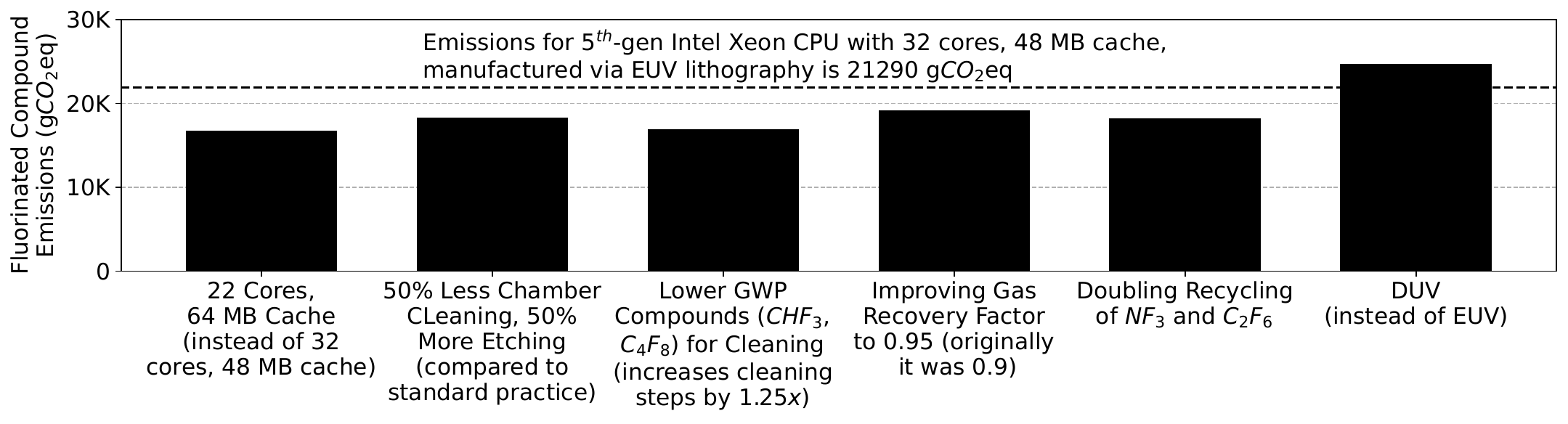}
    \vspace{-4mm}
    \caption{Fabrication facilities can use \sol{} to form design and manufacturing decisions for controlling fluorinated compound emissions.}
    \label{fig:interchange}
    \vspace{-4mm}
\end{figure*}

\subsection{Reduction of Fluorinated Compound Emissions by Using \sol{}}
\label{sec:interchange}

Fabrication facilities can utilize \sol{} to alter the specifications of their design, resource usage in different manufacturing steps, and manufacturing technology to minimize emissions of fluorinated compounds. For example, as shown in Fig.~\ref{fig:interchange}, 21290 gCO$_2$eq of fluorinated compound is emitted for manufacturing of a $5^{th}$-generation Intel Xeon CPU with 32 cores and 48 MB cache at Intel's Oregon fabrication facility, via EUV lithography. However, to reduce emissions, Intel can choose to reduce the number of cores, while increasing the cache size of the processor to still maintain similar performance in terms of the number of floating point operations per unit of time. Since decreasing core counts reduces the area of a die ($A_{\text{Die}}$) at a higher rate than increasing of $A_{\text{Die}}$ due to an increase in cache size, the overall $A_{\text{Die}}$ reduces, reducing the number of wafers ($N_{\text{Wafers}}$) required to manufacture a processor, which in turn reduces the overall emissions, as shown in Fig.~\ref{fig:interchange}. Reducing the number of cores from 32 to 22, while increasing the cache size from 48 MB to 64 MB, reduces fluorinated compound emissions by 22\%. 
However, both designs maintain similar performance (0.1 TFLOPs) because the larger cache enhances data locality, reducing memory bottlenecks and improving the efficiency of each core. Fewer cores free up thermal and power budgets, enabling higher clock speeds to offset reduced parallel processing.

Next, facilities can reduce emissions by optimizing complementary steps, such as decreasing the frequency of chamber cleaning while increasing etching efficiency, as reduced residue accumulation in optimized etching minimizes the need for cleaning. We observe from Fig.~\ref{fig:interchange}, that as we reduce chamber cleaning by 50\% and increase the etching steps by 50\%, the emissions of fluorinated compounds reduce by 14.5\%. This is because the GWP of the fluorinated compounds used for etching is much lower than the GWP of the compounds used for chamber cleaning (Table.~\ref{tab:model}).

Furthermore, fabrication facilities can choose to alter material usage to reduce emissions from the sources, like chamber cleaning, which contributes to significant emissions. Traditional chamber cleaning gases such as NF\textsubscript{3} and SF\textsubscript{6} have GWPs exceeding 17{,}000 and 25{,}000 times of CO\textsubscript{2} respectively. By contrast, CHF\textsubscript{3} and C\textsubscript{4}F\textsubscript{8} exhibit significantly lower GWPs (Fig.~\ref{fig:global_warming_potential}), which makes them attractive replacements that can shrink the emission~\cite{allgood2003fluorinated}. A practical trade-off, however, is that lower-GWP gases often require more frequent or longer cleaning cycles to achieve comparable residue removal in chamber interiors, thus our model increases the number of cleaning steps by $\approx$1.25$\times$ compared to standard practice. We derive this 1.25$\times$ factor by comparing the relative gas reactivity of CHF\textsubscript{3} and C\textsubscript{4}F\textsubscript{8} to that of NF\textsubscript{3} and SF\textsubscript{6}.

Recall from Eq.~\ref{eq:total}, that the net emissions depend directly on 
\((1 - \eta_{\text{Rec}})\), where \(\eta_{\text{Rec}}\) is the gas recovery factor. Increasing 
\(\eta_{\text{Rec}}\) from 0.90 to 0.95 means only 5\% of the fluorinated compounds escape as emissions, rather than 10\%. This improvement can be achieved through tighter containment, better abatement equipment, and more efficient vacuum pump capture. Even a modest bump in 
\(\eta_{\text{Rec}}\) has an outsized impact on lowering the final \(\text{gCO}_2\text{eq}\) of emission, because the total usage of fluorinated gases is typically large and the GWP of each molecule is extremely high. For example, increasing \(\eta_{\text{Rec}}\) from 0.90 to 0.95 for manufacturing of a $5^{th}$-generation Intel Xeon CPU at Intel's Oregon fabrication facility reduces fluorinated compound emissions by more than 10\% (Fig.~\ref{fig:interchange}).

Additionally, fabrication facilities can implement on-site gas-recovery lines and install closed-loop re-feeds to reclaim twice as much NF\textsubscript{3} and C\textsubscript{2}F\textsubscript{6} for reuse, rather than disposing them~\cite{schottler2008carbon,
dammel2024pfas}. Both of these gases are highly effective for etching and cleaning, so capturing them post-process and re-injecting them into the supply can lower net consumption and emissions. Using \sol{}, we calculated the effect by modeling half-equivalent emissions in the recovery process, thereby significantly reducing overall gCO\textsubscript{2}eq emissions (Fig.~\ref{fig:interchange}). 

Finally, different forms of manufacturing (\textit{e.g.,} EUV vs DUV photolithography) also impact the emissions, as shown in Fig.~\ref{fig:interchange}. Hence, fabrication facilities can use \sol{} to model emissions from different sources and make choices of hardware design and manufacturing steps to reduce emissions. Next, we discuss the methodology to determine the parameters used in \sol{}, enabling researchers to account for fabrication-specific practices and material usage.

\begin{figure*}[t]
    \centering
    \includegraphics[scale=0.58]{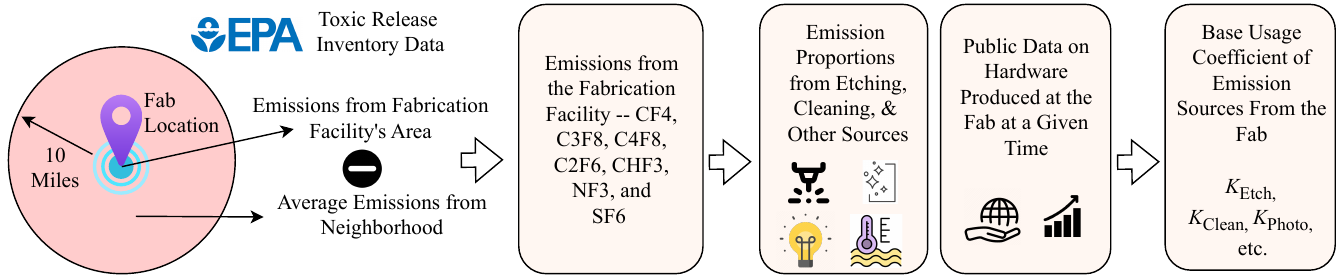}
    \vspace{-4mm}
    \caption{We develop a methodology to estimate the emission parameters of fabrication facilities using public toxic release datasets. This will help researchers to analyze emissions for manufacturing hardware at different fabrication facilities.}
    \vspace{-4mm}
    \label{fig:metho_cartoon}
\end{figure*}

\subsection{Methodology to determine fab-specific and reference hardware parameters}
\label{sec:methodology}

\noindent Fabrication-specific data is typically proprietary and not openly shared by facilities. We leverage fabrication process simulators, semiconductor supply chain datasets, published studies, and open-source emissions data to estimate parameters for emission sources, reference hardware, and facility practices, aligning our emission modeling with real-world manufacturing constraints.

To set parameters for the reference hardware and scaling factors in \sol{}'s fluorinated compound usage model (Table~\ref{tab:model}), we use FabSim~\cite{vogt2004new, fabsimFabSimSemiconductor}, a discrete-event simulator that covers over 300 fab process steps to optimize operations, tool use, and scheduling. By integrating FabSim with tools like LithographySimulator~\cite{gau2023ultra,githubGitHubQuarterwave0LithographySimulator}, we simulate modern lithography (e.g., EUV) to manage photolithography, etching, and related steps under realistic conditions. The Advanced Semiconductor Supply Chain Dataset~\cite{etoDocumentationAdvanced} provides essential details on tools, materials, and processes, letting us set accurate inputs for equipment use and emissions. The IEEE IRDS~\cite{ieeeirds} outlines fab steps and variations by node size (e.g., 7 nm, 14 nm), guiding our scaling and usage parameters across nodes. The SECOM Dataset~\cite{uciMachineLearning} helps identify emissions and yield factors through feature selection, while published works on semiconductor manufacturing~\cite{wu2014extreme,
wurm2020euv,geng2005semiconductor,khan2021semiconductor,moreau2012semiconductor,nojiri2015dry,quirk2001semiconductor} allow us to model steps like etching, deposition, oxidation, and photolithography in line with industry standards.

To estimate facility-specific emissions parameters, such as the base usage coefficients for different sources,  we use the Toxics Release Inventory (TRI) dataset~\cite{epaToolbox} from the United States Environmental Protection Agency~\cite{national2011sustainability}, and from European Chemical Agency~\cite{europaPerPolyfluoroalkyl} which provides data on environmental releases at specific geographic coordinates at different time frames. Since emissions at the facility’s geographical location may include contributions from other sources, we calculate an average emission value from the surrounding 10-mile radius as a baseline. By subtracting this baseline from the facility's reported emissions, we isolate the emissions that originate specifically from the fabrication processes. Thereafter, we use publicly available data on the number of chips manufactured every year~\cite{mckinseySemiconductorFabs,z2dataStatisticsSemiconductor,
wikipedia_fab}, the types of chips being manufactured by facilities in a timeframe~\cite{intelIntelProduct,githubGitHubToUpperCase78intelprocessors}, and the compounds used by each of the emission sources during manufacturing~\cite{interfaceeuChipProductions,czerniak2018pfc,
epaRiskManagement} to determine the base usage coefficients by each source. By estimating these parameters and incorporating the new hardware specifications to be manufactured, \sol{} enables researchers to model emissions for new hardware in advance, facilitating emission optimization from each source before finalizing the manufacturing pipeline design. This methodology, summarized in Fig.~\ref{fig:metho_cartoon}, bridges the gap between proprietary fabrication data and open-source research. Next, we evaluate the accuracy of \sol{}'s emission modeling.


\subsection{\sol{}'s modeling accuracy}
\label{sec:accuracy}
\noindent Here, we compare the effectiveness of \sol{}'s fluorinated compound emission modeling (Sec.~\ref{sec:model_em}) by comparing it with the emissions from fabrication facilities (determined via collecting the emissions value from the TRI dataset during the time when the hardware being evaluated was being manufactured, as described in Sec.~\ref{sec:methodology}). 

\begin{figure}[t]
    \centering
    \includegraphics[scale=0.36]{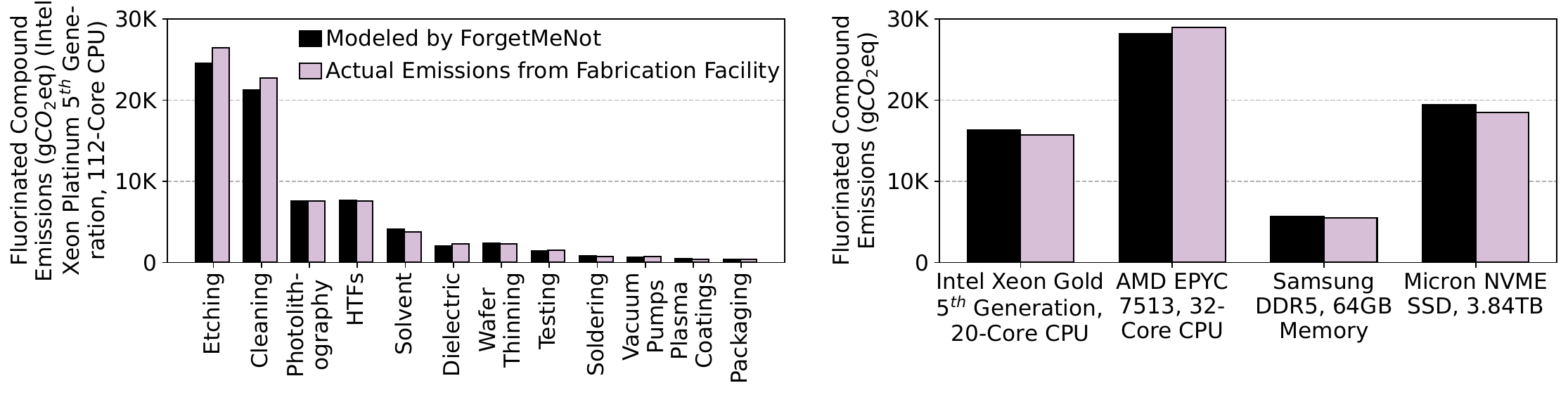}
    \vspace{-4mm}
    \caption{\sol{}'s modeling closely matches the emissions from each source from fabrication facilities (left). \sol{}'s modeling closely matches the total emissions from fabrication facilities for different hardware (right).}
    \label{fig:comparison_by_source}
    \vspace{-4mm}
\end{figure}

First, from Fig.~\ref{fig:comparison_by_source} (left), we note that the emissions from each source (as described in Sec.~\ref{sec:model_em}) as estimated by \sol{} closely match with the emissions from each source of the fabrication facility while manufacturing a $5^{th}$ generation Intel Xeon Platinum CPU with 123 cores (reference older generation hardware for modeling is as described Sec.~\ref{sec:model_em}). Additionally, from Fig.~\ref{fig:comparison_by_source} (right) we note that for different hardware components (Intel CPU manufactured in Oregon (USA), AMD CPU manufactured in Dresden (Germany), Samsung DRAM manufactured in Texas (USA), and Micron SSD manufactured in Idaho (USA)), the net emissions modeled by \sol{} closely matches the emissions from fabrication facility (a maximum difference of 4.7\%). This shows that \sol{} can effectively generalize modeling across different fabrication facilities and hardware types, accurately modeling emissions based on the process parameters and facility-specific practices. Next, we use \sol{} to analyze emission characteristics of manufacturing different hardware. 


\begin{myregbox}{yellow}{} 
\textbf{Takeaway:} \sol{} models emissions from key sources of fluorinated compounds by integrating fabrication-specific practices, reference older hardware parameters, and new hardware specifications. It enables fabs to estimate emissions for new hardware before designing the manufacturing pipeline, allowing analysis and optimization of emissions from each source to support sustainable manufacturing decisions. Furthermore, \sol{} enables facilities to explore trade-offs, such as adjusting hardware designs or optimizing steps like reducing chamber cleaning and increasing etching, to minimize emissions without compromising quality.
\end{myregbox}
\section{\sol{}: Analysis of Emissions}
\label{sec:analysis}

In this section, we use \sol{} to model and analyze the fluorinated compound emissions trends for manufacturing CPU, DRAM, and storage components. Additionally, we compare fluorinated compound emissions with the embodied carbon, which is the carbon emitted during the manufacturing of hardware. \sol{} uses a widely-used publicly available dataset~\cite{davy2021building, romain} to determine the embodied carbon of different hardware. This dataset is based on Dell's carbon footprint data~\cite{stutz2012carbon} and combines information from multiple sources~\cite{boyd2011life, auger2021open} to accurately determine the embodied carbon. We do not analyze the operational carbon of hardware devices. Prior works have analyzed it and it depends on the amount of hardware usage (results in energy consumption) and location of hardware (carbon intensity varies with location)~\cite{lee2024carbon,
maji2024untangling}. Combining embodied carbon and fluorinated compound emissions results in the total emissions during manufacturing. 

\vspace{2mm}

\begin{figure}[t]
    \centering
    \includegraphics[scale=0.37]{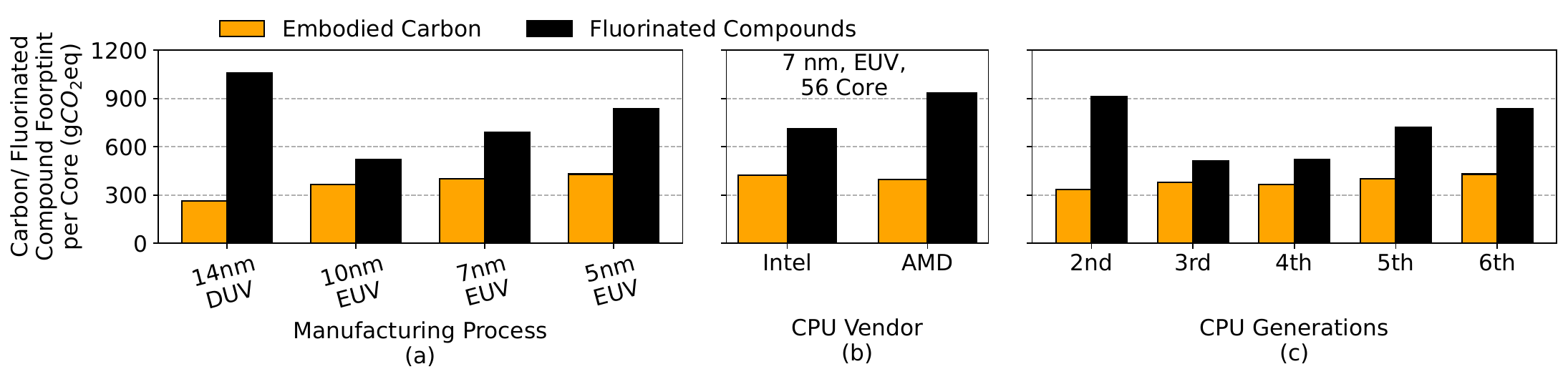}
    \vspace{-4mm}
    \caption{Emissions vary across used manufacturing technologies, vendors, and hardware generations.} 
    \label{fig:cpu}
    \vspace{-4mm}
\end{figure}



\noindent\textbf{Emissions from CPU manufacturing.} From Fig.~\ref{fig:cpu}, we observe that the fluorinated compound footprint remains consistently higher than the embodied carbon footprint throughout the CPU manufacturing process. For example, the fluorinated compound emissions are 3.02$\times$ higher than the embodied carbon when manufacturing a 14 nm CPU at Intel (Oregon) fabrication facility using DUV (Fig.~\ref{fig:cpu}(a)). Despite being used in small quantities, the extremely high GWPs and persistent nature of fluorinated compounds result in emissions that surpass that of embodied carbon emissions in CPU manufacturing. Also, several factors, including the CPU’s manufacturing processes, lithography techniques, vendor/ fabrication-specific practices, and hardware generations influence emissions. 

In Fig.~\ref{fig:cpu}(a), we show 4 types of Intel CPUs under different nanometer technology manufacturing processes and lithography techniques (DUV and EUV). Each type reveals distinctive environmental impacts based on the process and technology used. For instance, DUV-based CPU (14 nm) tends to result in higher fluorinated compound emissions compared to those manufactured with EUV lithography at smaller nodes (5, 7, 10 nm). This difference is largely due to EUV’s higher efficiency in patterning extremely small transistors ($\phi_{\text{Lith}}$ in Table~\ref{tab:model}), thereby reducing the need for additional fluorinated compounds. Furthermore, as the manufacturing node size (nanometer technology) decreases, both carbon and fluorinated compound footprints increase. In Fig.~\ref{fig:cpu}(a), we illustrate this trend by modeling emissions as the node size decreases from 10 nm to 7 nm and then to 5 nm. Compared to the original 10 nm Intel EUV CPU, the embodied carbon emissions increase by 9.6\% when moving to 7 nm and by a further 17.8\% at 5 nm. Similarly, the fluorinated compound emissions rise by 17.5\% for 7 nm and by an additional 26.9\% for 5 nm.

Next, we apply \sol{} to analyze emissions from CPUs manufactured by different vendors -- Intel and AMD CPUs with 56 cores, both manufactured using a 7 nm process via EUV lithography (Fig.~\ref{fig:cpu}(b)). Intel CPU tends to have 7.5\% more carbon compound emissions, compared with AMD CPU. However, AMD CPU generated more fluorinated compounds during manufacturing; the fluorinated compound footprint of AMD CPU is 31.7\% higher than Intel. This is because, Intel's manufacturing pipeline is highly optimized with a lesser number of etching, lithography, and deposition steps and lower fluorinated compound usage per manufacturing step due to Intel's in-house fabrication facilities, compared to fabrication outsourcing to TSMC by AMD. 

\begin{figure}[t]
    \centering
    \includegraphics[scale=0.36]{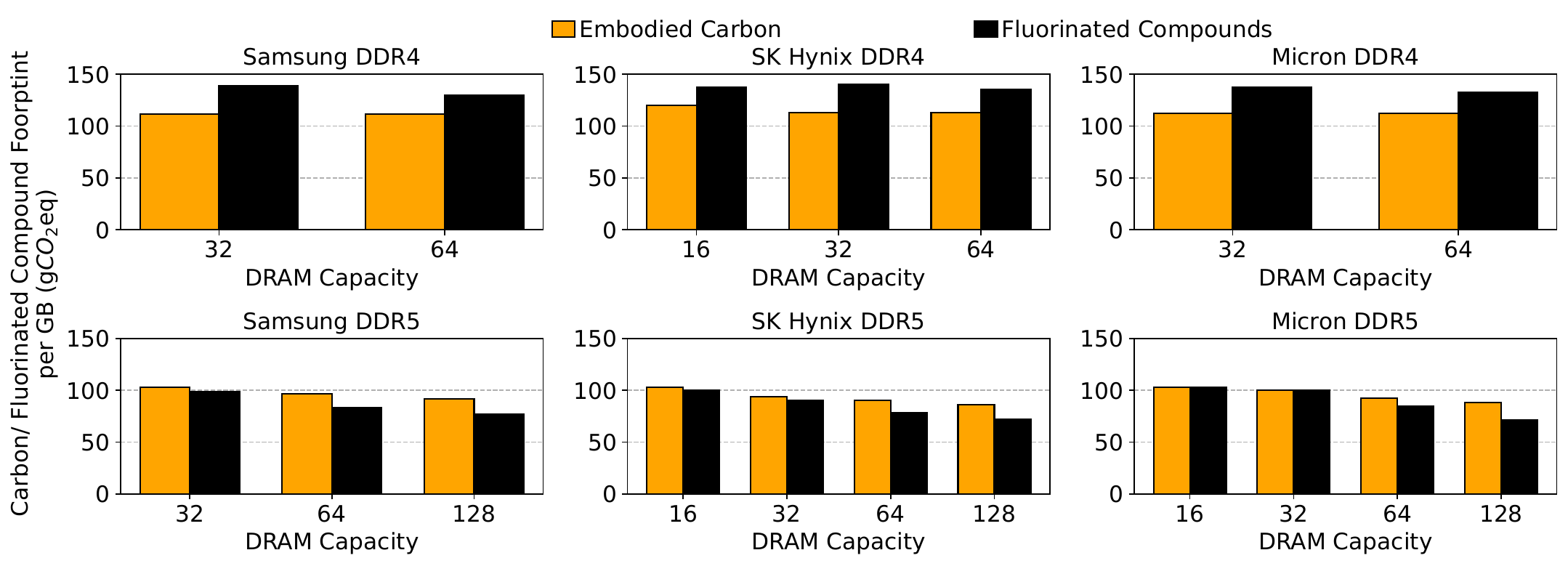}
    \vspace{-4mm}
    \caption{DDR5 memory modules have lower environmental footprints than DDR4 memory modules, especially at higher DRAM memory capacities.}
    \label{fig:dram}
    \vspace{-4mm}
\end{figure}

Finally, we use \sol{} to analyze multiple generations of Intel Xeon Platinum CPUs (Fig.~\ref{fig:cpu}(c)). The results reveal that newer generations of Intel CPUs lead to higher embodied carbon emissions due to the adoption of advanced manufacturing techniques and materials, which, despite enhancing performance, involve more energy-intensive processes. In terms of fluorinated compounds, the analysis shows a significant increasing trend in their footprint with each generation, beyond the 2$^{nd}$ generation CPU. The drop in fluorinated compound emissions from 2$^{nd}$ to 3$^{rd}$ generation CPU is due to changing from DUV to EUV lithography. This trend highlights the growing environmental impact associated with the manufacturing of more advanced Intel processors. Each new CPU generation introduces smaller node sizes, requiring more manufacturing steps resulting in the usage of more fluorinated compounds, thus increasing emissions. 

\vspace{2mm}



\noindent\textbf{Emissions from DRAM manufacturing.} In Fig.~\ref{fig:dram}, we show a comparison, in terms of carbon and fluorinated compound footprints, for different types of DRAM, specifically SK Hynix, Samsung, and Micron DDR4 and DDR5 memory across various capacities. It demonstrates that both carbon and fluorinated compound emissions per GB of memory remain almost the same as the size of the entire memory module increases for DDR4 memory, and only slightly decreases for DDR5 memory, due to more efficient use of materials and reduction of manufacturing steps. Additionally, DDR5 modules have lower footprints compared to their DDR4 counterparts of the same memory module capacity due to more streamlined fabrication processes and reduced material usage enabled by higher memory density. Thus newer DDR5 DRAM modules are efficient in terms of embodied carbon footprint, and fluorinated compound emissions. Unlike CPU manufacturing, fluorinated compound emissions are not very different from embodied carbon emissions in DRAM. Also, for server-grade nodes with more than 20 cores, and 32 - 64 GB memory, the impact of emissions from memory modules is much lower than that from CPU manufacturing. 

\begin{figure}[t]
    \centering
    \includegraphics[scale=0.35]{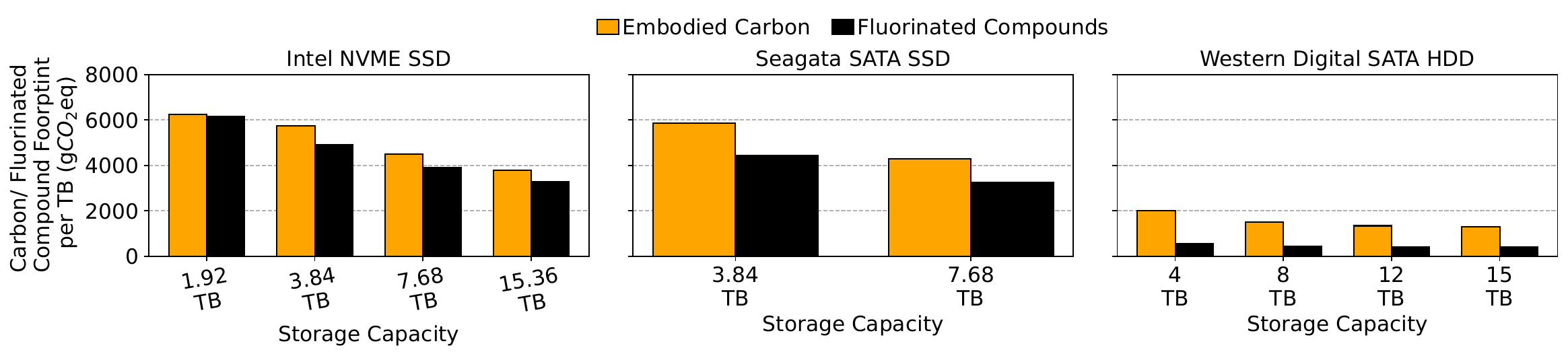}
    \vspace{-4mm}
    \caption{Carbon and fluorinated compound footprints per TB reduce with increased storage capacity. HDDs have a significantly lower environmental footprint than SSDs.}
    \label{fig:storage}
    \vspace{-4mm}
\end{figure}


\vspace{2mm}

\noindent\textbf{Emissions from storage manufacturing.} From Fig.~\ref{fig:storage}, we observe that there is a significant reduction in the carbon and fluorinated footprints per terabyte for different storage types as the storage capacity increases as the number of manufacturing steps only increases sub-linearly with an increase in capacity. For Intel NVME SSDs, the carbon and fluorinated compound footprints of 15.36 TB both decrease nearly 40\% compared to those at 1.92 TB. Similarly, Seagate SATA SSDs demonstrate a 35\% reduction in embodied carbon footprint and a 25\% reduction in fluorinated compound footprint as the storage capacity scales up from 4 TB to 8 TB. This trend emphasizes the environmental efficiency of adopting higher-capacity storage solutions, driven by economies of scale and advancements in manufacturing processes.

Although different storage modules follow similar emission trends, NVMe SSDs and SATA SSDs exhibit notable differences. NVMe SSDs are generally faster and consume more power, while SATA SSDs are a more budget-friendly option for upgrading storage in standard systems. From Fig.~\ref{fig:storage}, we observe that the embodied carbon footprints of a 3.84 TB Intel NVMe SSD and a 3.84 TB Seagate SATA SSD differ by less than 3\%. However, the fluorinated compound footprint of the SATA SSD is 10\% lower than that of the NVMe SSD. Note that, unlike DRAM, the environmental footprint of storage is significantly large and can contribute to a major emission source when designing a complete server. From Fig.~\ref{fig:storage}, we observe that HDDs have a significantly lower environmental footprint than SSDs. For example, using a 4TB Western Digital SATA HDD can save 66\% and 87\% carbon and fluorinated compounds compared to using a 3.84 TB Seagate SATA SSD. The fluorinated compound footprints follow a similar pattern, with HDDs consistently demonstrating lower values across all capacities. Despite the performance trade-offs, this underscores SATA HDDs as a more sustainable storage option than SSDs, particularly for large-scale deployments. 

Next, we use this analysis and \sol{}'s modeling to assemble components for designing sustainable datacenter servers.

\begin{myregbox}{yellow}{}
\textbf{Takeaway:} CPU manufacturing exhibits significantly higher fluorinated compound emissions than embodied carbon, influenced by manufacturing techniques, lithography, and hardware generations. For DRAM, DDR5 modules have lower environmental footprints than DDR4 due to improved fabrication efficiencies. SSDs form a major source of fluorinated compounds and carbon emissions in servers. Storage solutions like HDDs offer reductions in both carbon and fluorinated emissions, highlighting the trade-off between sustainability and performance.
\end{myregbox}

\section{Using \sol{} to Build Environmentally Sustainable Servers}
\label{sec:server}

\begin{table*}[t]
\centering
\caption{Providers can choose between different hardware components that result in lowering total emissions (embodied carbon and fluorinated compounds) during the manufacturing of servers.}
\label{tab:opt}
\vspace{-2mm}
\renewcommand{\arraystretch}{1.5}
\setlength{\tabcolsep}{4pt}
\footnotesize{
\begin{tabular}{|p{0.15\linewidth} | p{0.185\linewidth} | p{0.185\linewidth} | p{0.185\linewidth} | p{0.19\linewidth}|}
\hline
\textbf{} & \textbf{General Purpose} & \textbf{Compute Optimized} & \textbf{Memory Optimized} & \textbf{Storage Optimized} \\ \hline
\cellcolor{red!30}\textbf{Highest Emission} & \cellcolor{red!30}175019 gCO$_2$eq & \cellcolor{red!30}175019 gCO$_2$eq & \cellcolor{red!30}175019 gCO$_2$eq & \cellcolor{red!30}175019 gCO$_2$eq \\ \hline
\cellcolor{orange!25}\textbf{Median Emission} & \cellcolor{orange!25}96741 gCO$_2$eq & \cellcolor{orange!25}98189 gCO$_2$eq & \cellcolor{orange!25}102389 gCO$_2$eq & \cellcolor{orange!25}114725 gCO$_2$eq \\ \hline
\cellcolor{green!20}\textbf{Lowest Emission} & \cellcolor{green!20}43300 gCO$_2$eq & \cellcolor{green!20}44360 gCO$_2$eq & \cellcolor{green!20}49260 gCO$_2$eq & \cellcolor{green!20}83300 gCO$_2$eq \\ \hline
\cellcolor{gray!10}\textbf{Lowest Emission Hardware} & \cellcolor{gray!10}$5^{th}$-gen Intel Xeon Gold with 20 cores, SK Hynix 32 GB DDR5 DRAM, Seagate 8TB HDD & \cellcolor{gray!10}$4^{th}$-gen Intel Xeon Gold with 24 cores, SK Hynix 32 GB DDR5 DRAM, Seagate 8TB HDD & \cellcolor{gray!10}$4^{th}$-gen Intel Xeon Gold with 24 cores, SK Hynix 64GB DDR5 DRAM, Seagate 8TB HDD & \cellcolor{gray!10}$5^{th}$-gen Intel Xeon Gold with 20 cores, SK Hynix 32GB DDR5 DRAM, Samsung 7.68TB NVMe SSD \\ \hline
\end{tabular}}
\vspace{-4mm}
\end{table*}

Sec.~\ref{sec:analysis} shows that the total emissions (embodied carbon and fluorinated compound) vary with hardware capacity, generations, vendors, and type. A service provider has the opportunity to reduce the environmental impact by procuring hardware components that meet performance goals while reducing emissions during manufacturing. If datacenter operators used hardware that emits lower emissions during manufacturing, then it would influence the demand for specific components, leading fabs to prioritize the production of lower-emission hardware while reducing the manufacturing of components with higher emissions. This market-driven approach could indirectly drive optimization in manufacturing practices to produce sustainable hardware.

\vspace{2mm}

\noindent\textbf{Building low-emission servers.} Table~\ref{tab:opt} demonstrates that cloud providers can intelligently assemble hardware components (CPU, memory, and storage) to build servers of different types, such as general purpose, compute optimized, memory optimized, and storage optimized while minimizing total emissions. For example, here we show an example of assembling mid-sized servers with 20 to 32 cores. We can select among various Intel Xeon Gold/Platinum cores between $2^{\text{nd}}$ to $5^{\text{th}}$ generations, or AMD EPYC cores. Node sizes (nanometer technology) vary between 14 nm to 7 nm, and cores are manufactured by EUV or DUV lithography. Memory options include various DDR4 and DDR5 memory of Samsung, SK Hynix, and Micron, with capacities between 32 to 128 GB. For storage, the possible options include various NVMe SSDs, SATA SSDs, and SATA HDDs between 7 TB to 16 TB capacity. To build an overall general-purpose instance, we have a total of 3300 choices, considering different combinations of CPU, memory, and storage to build a server. The lowest emission hardware combination reduces manufacturing emissions by 75\% compared to the highest-emission setup and by 55\% compared to the median-emission configuration. We note that the lowest emission hardware uses HDDs as they have significantly lower emissions than SSDs, and uses CPUs with 20 cores (even though CPUs up to 32 cores are available in the possible choices), as emissions scale up with an increase in cores. 

Similarly, we observe that among compute optimized instances (instances with core counts greater than 24 cores, and belonging among newer ($4^{\text{th}}$ or  $5^{\text{th}}$) generations, as they are typically faster) and memory optimized instances (instances with core counts between 20 - 32 cores, and DDR5 memory of greater than 32 GB capacity as they are faster than DDR4 alternatives), the instances with lowest emission reduces total emission by 54\% and 51\%, respectively, over the instances with median total emissions. There are a total of 2970 compute optimized server choices and 1080 memory optimized server choices to choose from. Note that, both compute and memory optimized instances use HDD for storage as they typically have much lower emissions than SSDs, which contribute toward lowering the total emissions of the assembled server.  

\begin{figure*}[t]
    \centering
    \includegraphics[scale=0.37]{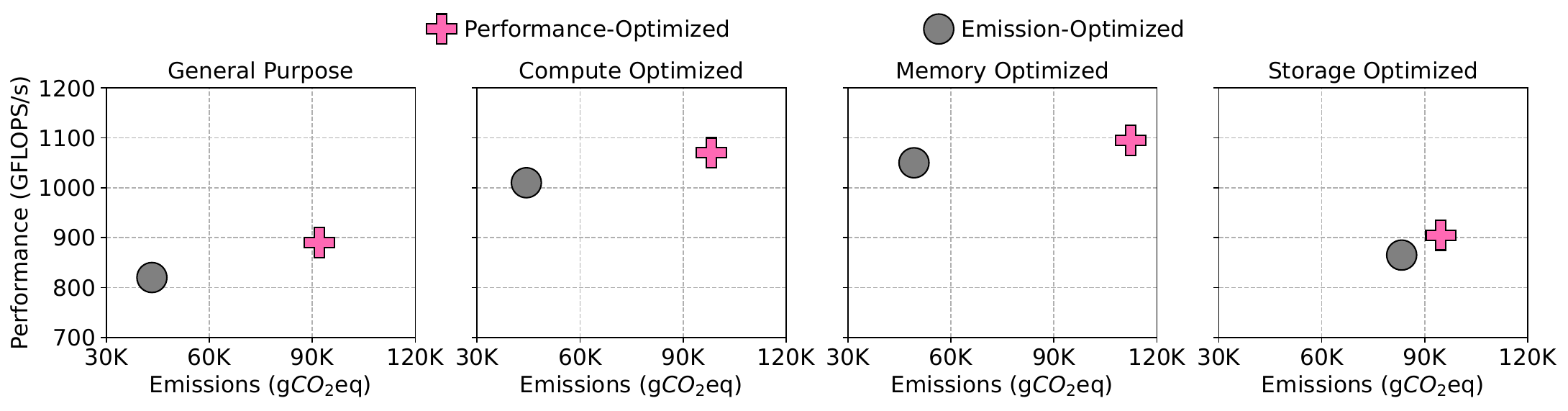}
    \vspace{-4mm}
    \caption{Performance-optimized configurations achieve only limited improvement in throughput but result in significantly higher emissions (summation of embodied carbon and fluorinated compound emissions) compared to emission-optimized configurations.}
    \vspace{-4mm}
    \label{fig:tradeoff}
\end{figure*}

Finally, we note that among storage optimized instances (instances with core counts between 20 to 32 cores and NVMe SSD storage, as they typically provide higher storage bandwidth compared to SATA SSDs and HDDs), the server assembly with the lowest emission, reduces total emission by 27\% compared to the server assembly with the median emissions among all choices of storage optimized server assembly (a total of 1980 choices). The difference between the server assembly with the median and the lowest emission is lower in this case compared to the other kinds of instances. This is because, for storage optimized instances all the server assembly choices use NVMe SSDs, which have higher emissions compared to other storage types like HDDs. Since all options rely on NVMe SSDs, the variation in emissions due to storage type is minimized, reducing the gap between emissions from different assembly configurations.

\todo{All emission estimates in Table~\ref{tab:opt} are based on a 500-year Global Warming Potential (GWP-500), which smooths the impact of long-lived gases over a longer time horizon. As discussed in Sec.\ref{sec:fluorine_motive}, GWP is time-dependent. Shorter time horizons such as GWP-100 or GWP-20 weigh short-lived but highly potent gases more heavily. Despite this, we find that the lowest emission hardware choices in Table~\ref{tab:opt} remain unchanged across GWP-500, GWP-100, and GWP-20 for general purpose, compute optimized, and memory optimized server types. For storage optimized servers, the lowest emission configuration under GWP-500 becomes the tenth-lowest under GWP-20 (out of a total of 1980 choices), due to minor shifts in the relative ranking of fluorinated compounds at shorter time horizons. Overall, this stability suggests that the emission optimized configurations are robust across standard GWP variants, as the relative ordering of fluorinated compound impacts does not vary significantly between GWP-500, GWP-100, and GWP-20~\cite{harvey1993guide}.}

\vspace{2mm}

\noindent\textbf{Balancing between performance and emissions.} While the configurations in Table~\ref{tab:opt} highlight the ones that minimize the manufacturing emissions, these low-emission choices may not yield the highest performance, which is of importance to datacenter operators to meet a certain level of quality of service. In practice, the fastest, or \textit{performance-optimized} instances in each category (general purpose, compute optimized, memory optimized, and storage optimized) tend to have maximum CPU cores (5th-generation Intel Xeon Platinum CPU manufactured using 7 nm EUV lithography), 128 GB DDR5 capacities, and high-performance NVMe SSDs. However, the performance optimized instances result in a significantly higher environmental cost due to being more manufacturing emission-heavy hardware.

For example, from Fig.~\ref{fig:tradeoff}, we observe that a general purpose, performance-optimized instance shows an 8.5\% increase in throughput compared to the lowest-emission (\textit{emission-optimized}) instance for Linpack HPL benchmarks, which are commonly used for ranking servers based on their number of floating point operation-wise performance~\cite{dongarra2003linpack}. However, the performance-optimized instance increases emissions by 52\%. In Fig~\ref{fig:tradeoff}, we consider the total manufacturing emissions: fluorinated compound and embodied carbon emissions. We do not report the operational emission as it can vary based on the utilization of hardware, its usage lifetime, and location of usage. However, the difference in the percentage of emissions between performance-optimized and emission-optimized hardware varies by less than 7\% when operational emissions are considered assuming 80\% server utilization for a 5-year period, and a 350 g$CO_2$eq/KWh carbon intensity (average in the USA)~\cite{electricitymap}. Similar trends between performance- and emission-optimized instances are observed for compute and memory optimized instances. The trade-off is less pronounced for storage-optimized instances, where the difference in emissions between the performance-optimized and emission-optimized configurations is 12\%. This is because all storage-optimized instances rely on NVMe SSDs, which are themselves emission-intensive components and dominate the overall emissions profile of the server. Overall, we observe a marginal performance gain for performance-optimized instances at the cost of substantially higher emissions, primarily because advanced node designs, larger memory modules, and high-capacity SSDs require more complex fabrication steps and materials that significantly increase fluorinated compound usage. Thus, when feasible, datacenter operators should select hardware configurations that more closely align with emission-optimized options, while still meeting the necessary quality of service targets.

\begin{table*}[t]
\centering
\caption{\todo{Ranking of five different 64-core CPUs based on embodied carbon, total emissions (embodied carbon footprint and fluorinated compound emissions), and performance in terms of TFLOPs (a lower ranking number is better).}}
\label{tab:cpu_rankings}
\vspace{-2mm}
\renewcommand{\arraystretch}{1.5}
\setlength{\tabcolsep}{4pt}
\footnotesize{
\begin{tabular}{|p{0.215\linewidth} | p{0.12\linewidth} | p{0.14\linewidth} | p{0.13\linewidth} | p{0.14\linewidth} | p{0.12\linewidth}|}
\hline
\textbf{Metric} & \textbf{AMD EPYC 7713} & \textbf{Intel Xeon Platinum 8481Y} & \textbf{AMD EPYC 7543} & \textbf{Intel Xeon Platinum 8492Y} & \textbf{AMD EPYC 7551} \\ \hline
\textbf{Ranking based on embodied carbon} & 2 & 1 & 4 & 3 & 5 \\ \hline
\textbf{Ranking based on total $CO_2-eq$ emission (embodied carbon and fluorinated compound)} & 4 & 2 & 5 & 1 & 3 \\ \hline
\textbf{Ranking based on performance (TFLOPs)} & 1 & 4 & 3 & 5 & 2 \\ \hline
\end{tabular}}
\vspace{-4mm}
\end{table*}

\todo{Finally, Table~\ref{tab:cpu_rankings} illustrates how the ranking of five different 64-core CPUs (each with 32 GB SK Hynix DDR5 memory, and Seagate 8TB HDD storage) can vary significantly depending on whether one optimizes for embodied carbon alone, total emissions (including both embodied carbon and fluorinated compounds), or overall performance (TFLOPs). For instance, while AMD EPYC 7713 leads in performance, Intel Xeon Platinum 8481Y has the lowest embodied carbon, and Intel Xeon Platinum 8492Y achieves the lowest total emissions when accounting for both embodied carbon and fluorinated compounds. As a result, current sustainability models commonly used in systems research, such as ACT~\cite{gupta2022act} which focus solely on embodied carbon (primarily CO$_2$), may overlook configurations that are more environmentally optimal when total emissions are considered. By combining \sol{}, which captures fluorinated compound emissions, with tools like ACT that estimate embodied carbon, cloud providers can achieve a more comprehensive and accurate assessment of hardware sustainability, enabling more informed and environmentally responsible procurement decisions. Additionally, this reinforces the point made in Fig.~\ref{fig:tradeoff}: performance-optimized configurations differ from emission-optimized ones, and \sol{} makes this trade-off explicit by providing critical visibility into the environmental costs of performance-centric hardware choices.}


\begin{myregbox}{yellow}{}
\textbf{Takeaway:} Datacenter operators can lower environmental impact by assembling servers with hardware optimized for reduced manufacturing emissions. General purpose, compute optimized, and memory optimized instances can cut emissions by up to 55\%, 54\%, and 51\%, respectively, compared to median emission setups. Storage optimized instances show smaller reductions (27\%) due to the higher emissions of NVMe SSDs. Additionally, while performance-optimized servers offer marginally higher throughput, they often incur much higher emissions, emphasizing the importance of balancing performance needs with sustainability goals. \todo{It is important to consider not just embodied carbon (primarily CO$_2$), but also emissions from fluorinated compounds, which can contribute significantly to the total manufacturing emissions.} Procuring low-emission hardware drives demand for sustainable components, encouraging fabs to innovate in low-emission manufacturing.
\end{myregbox}

\section{Related Works}
\label{sec:rel}

\noindent\textbf{Sustainability-centric frameworks in computing.}
Environmental sustainability has become an important consideration in computer system design, with key indicators being carbon emissions~\cite{gupta2022act,wang2024designing,li2023toward,gupta2021chasing,acun2023carbon,eeckhout2024focal,hanafy2024going,jiang2024ecolife}, and water usage~\cite{li2023making, gnibga2024flexcooldc}. While PFAS in semiconductor manufacturing has been highlighted~\cite{elgamalenvironmental}, most efforts to date address carbon emissions, particularly operational carbon, through techniques like scheduling optimizations and server utilization improvements~\cite{sukprasert2024limitations,
hanafy2024going, bashir2024sunk,jiang2024ecolife}. Embodied-carbon research has expanded as well. For instance, a Kaya-inspired analysis~\cite{eeckhout2023kaya} projects that embodied emissions could soon outstrip operational emissions, prompting an increased focus on chip design. Works like GreenSKU~\cite{wang2024greensku} focus on optimizing the carbon efficiency of cloud computing servers by designing carbon-efficient server units using low-carbon components, such as energy-efficient CPUs and reused DRAM and SSDs. It focuses on balancing operational and embodied emissions, offering detailed, server-level optimizations that complement ACT’s~\cite{gupta2022act} broader, architectural-level embodied carbon modeling framework. ECO-CHIP~\cite{sudarshan2024ecochip} on the other hand, targets embodied carbon emissions in chiplet-based heterogeneous architectures by integrating advanced packaging techniques and hierarchical design approaches. Complementing these works, \sol{} models the overlooked fluorinated compounds, integrating fab-specific chemical usage, lithography details, and hardware attributes to offer a more broader view of embodied emissions.

\vspace{2mm}

\noindent\textbf{Fluorinated compound emission reductions in other industries.} Multiple sectors have developed strategies to reduce high-GWP fluorinated gases, particularly SF\textsubscript{6}. For instance, the power industry has introduced natural or synthetic blends for switchgear applications~\cite{koch2003sf6,franck2021sf6free,burges2020sf6}, magnesium production cut SF\textsubscript{6} usage by 50\% through phase-out initiatives~\cite{epa2013global}, and the aluminum sector shifted to alternative gases or mechanical methods to lower degassing emissions~\cite{iai2019perfluorocarbon}. LCD manufacturing has likewise reduced SF\textsubscript{6} usage through improved process control and gas mixtures~\cite{cheng2013sf6usagelcd}. Yet, these measures typically target a narrower set of fluorinated compounds and contexts than those found in semiconductor fabrication, which involves a variety of fluorinated compound usage across etching, lithography, cleaning, and other steps. No existing framework systematically addresses the breadth of these emissions in hardware manufacturing, a gap that \sol{} aims to fill.

\section{Conclusion}
This paper establishes the importance of accounting for fluorinated compounds, which are more potent than carbon dioxide in terms of global warming potential, in computing hardware manufacturing. By introducing \sol{}, we provide a modeling framework that incorporates fabrication-specific practices, and hardware attributes to comprehensively estimate forever chemical emissions, aiding fabrication facilities and datacenter operators in building sustainable systems. We hope this work motivates researchers and designers to address the environmental impact of forever chemicals, promoting a broader and more sustainable approach to computing.

\vspace{2mm}

\noindent\todo{\textbf{Acknowledgment.} We thank David Irwin (our shepherd) and the reviewers for their constructive feedback. This work is supported by NSF Awards 1910601, and  2124897.}


\balance
\bibliographystyle{plain}
\bibliography{refs}

\end{document}